\documentclass[smallextended]{svjour3}       
\smartqed  
\usepackage{graphicx}

\usepackage{mathptmx}   
\usepackage[]{algorithm2e}   
\usepackage{graphics} 
\usepackage{epsfig} 
\usepackage{mathptmx} 
\usepackage{times} 
\usepackage{amsmath} 
\usepackage{amssymb}  

\usepackage{amsthm}
\usepackage{graphicx}
\usepackage{epstopdf}
\usepackage{comment}
\usepackage{color}
\usepackage{cite}
\usepackage{xcolor}
\usepackage{graphicx}
\usepackage{float}
\usepackage{epstopdf}
\usepackage[]{algorithmic}
\usepackage[justification=centering]{caption}

\floatname{algorithm}{Procedure}

\begin{document}

\title{Observation-Assisted Heuristic Synthesis of Covert Attackers Against Unknown Supervisors
%
}


\author{
Liyong Lin, Ruochen Tai, Yuting Zhu, Rong Su
}

\authorrunning{L. Lin, R. Tai, Y. Zhu, R. Su} 

\institute{Liyong Lin, Ruochen Tai, Yuting Zhu, Rong Su \at
              Electrical and Electronic Engineering, Nanyang Technological University, Singapore \\
\email{liyong.lin@ntu.edu.sg, ruochen001@e.ntu.edu.sg, yuting002@e.ntu.edu.sg, rsu@ntu.edu.sg} 
          }


\maketitle

\begin{abstract}
In this work, we address the problem of  synthesis of covert  attackers in the  setup where the model of the plant is available, but the model of the supervisor is unknown, to the adversary. To compensate the lack of knowledge on the supervisor, we assume that the adversary has recorded a (prefix-closed)  finite set of observations of the runs of the closed-loop system, which can be used for assisting the synthesis.  We present a  heuristic algorithm for the synthesis of covert damage-reachable attackers, based on the model of the plant and the (finite) set of observations,
by a transformation into solving an instance of the partial-observation supervisor synthesis problem.  The heuristic algorithm developed in this paper may allow the adversary to synthesize covert  attackers  without having to know the model of the supervisor, which could be  hard to obtain in practice. For simplicity, we shall only consider covert attackers that are  able to  carry out sensor replacement attacks and actuator disablement attacks. The effectiveness of our approach is illustrated on a water tank example adapted from the literature. 
\end{abstract}

\section{Introduction}

The security of cyber-physical systems, modelled in the abstraction level of events~\cite{WMW10}, has  drawn much research interest from  the discrete-event systems  community, with most of the existing works devoted to  attack detection and  security verification~\cite{CarvalhoEnablementAttacks, Carvalho2018, LACM17, Lima2018, WTH17, WP},  synthesis of covert attackers~\cite{ Su2018, Goes2017, Goes2020, LZS19, Lin2018, LS20, Kh19, LS20J, Mohajerani20} and  synthesis of resilient supervisors~\cite{Su2018, GSS19, LZS19b, Zhu2018, WBP19, Su20, LS20BJ}. Intuitively, the covertness property says that the attacker cannot reach a situation where its presence has been detected by the monitor but no damage can be inflicted~\cite{LS20, LS20J}. Thus, the covertness property is a safety property for the attacker. In this paper, we focus on the synthesis of covert attackers in a more practical setup than those of~\cite{Su2018, Goes2017, Mohajerani20, Goes2020, LZS19, Lin2018, LS20, Kh19, LS20J}. 

The problem of covert sensor attacker synthesis has been studied extensively~\cite{Su2018, Goes2017,Mohajerani20, Goes2020}. In~\cite{Su2018}, it is shown that,
under a normality assumption on the sensor attackers, the supremal covert sensor attacker  exists and can be effectively synthesized. In~\cite{Goes2017, Goes2020}, a game-theoretic approach is presented to synthesize covert sensor attackers, without imposing the normality assumption. Recently, based on the game arena  of~\cite{Goes2017, Goes2020}, ~\cite{Mohajerani20} develops an  abstraction based synthesis approach to improve the synthesis efficiency of~\cite{Goes2017, Goes2020}. The problem of covert actuator attacker synthesis has been addressed in~\cite{Lin2018} and~\cite{LZS19}, by  employing a reduction to the (partial-observation) supervisor synthesis problem~\cite{LZS19}. With the reduction based approach,  the more general problem of covert actuator and sensor attacker synthesis has also been addressed~\cite{Kh19,LS20J, LS20}. 

The synthesis approach developed in the existing works is quite powerful, in the sense that maximally permissive covert attackers can be synthesized from the model of the plant and the model of the supervisor.   While it is  natural to assume the model of the plant to be known, it seems a bit restrictive  to also assume the model of the supervisor to be  known, which could limit the usefulness of the existing covert attacker synthesis procedures in practice. In this paper, we relax the assumption that the model of the supervisor is known. We  assume  the adversary has recorded a (prefix-closed) finite set of observations of the runs of the closed-loop system, which can be used for assisting the synthesis of covert attackers. In this new setup, a covert attacker needs to be synthesized, if it is possible, based (solely) on the model of the plant and the given set of observations. 
The synthesized attacker needs to ensure\footnote{From the adversary's point of view, any supervisor that is consistent with the given set of observations may have been deployed.} damage-infliction and covertness against all the supervisors which are consistent with the given set of observations. The difficulty of this synthesis problem lies in the fact that there is in general an infinite number of supervisors that are consistent with the observations, rendering the synthesis approach developed in the existing works ineffective. 

In this work, we consider  covert  attackers whose attack mechanisms are  restricted to sensor replacement attacks and actuator disablement attacks\footnote{It is worth mentioning that sensor replacement attacks can help achieve both of the damage-infliction goal and the covertness goal; actuator disablement attack can help achieve the covertness goal.}. For simplicity, we only address the problem of synthesis of  covert damage-reachable attackers~\cite{LS20, LS20J}.  The main contributions of this work are listed as follows.
\begin{enumerate}
    \item [$\bullet$] We consider a new, but more challenging, setup where  covert  attackers  need to be synthesized solely based on the model of the plant and a (prefix-closed) finite set of observations of the runs of the closed-loop system. This effectively removes the assumption that the model of the supervisor is known (a prior) to the adversary, which is assumed in~\cite{ Su2018, Goes2017, Goes2020, LZS19, Lin2018, LS20, Kh19, LS20J, Mohajerani20} that address the covert  attacker synthesis problem. 
    \item [$\bullet$] We provide a  heuristic algorithm for the synthesis of covert  attacker, based solely on the model of the plant and the given set of observations. The solution methodology is to formulate the covert  attacker synthesis problem in this new setup as an instance of  the  (partial-observation) supervisor synthesis problem, and  it follows that we can  employ the existing supervisor synthesis solvers~\cite{Feng06, Susyna, Malik07} to synthesize covert attackers, without knowing the model of the supervisor. 
    The effectiveness of our approach is illustrated on a  water tank example adapted from~\cite{Su2018}.
    \item [$\bullet$] We explain how the correctness of the synthesis approach can be reasoned based on the model of the attacked closed-loop system adapted from~\cite{LS20},~\cite{LS20J},~\cite{LS20BJ}.
\end{enumerate}
This paper is organized as follows. In Section 2, we recall the preliminaries which are needed for better understanding this work. In Section 3, we  then introduce the system setup, present the model constructions, the proposed synthesis solution as well as the correctness proof. To establish the correctness of the  synthesis solution, we recall the model of the attacked closed-loop system, which is adapted from~\cite{LS20},~\cite{LS20J},~\cite{LS20BJ}, in Section 3. We also provide a brief discussion on the time complexity of our heuristic synthesis algorithm in Section 3.
Finally, in Section 4, conclusions and future works are discussed.  

\section{Preliminaries}
In this section, we introduce some basic notations and terminologies that will be used in this work, mostly following~\cite{WMW10, CL99, HU79}.  
 
 
 For any two sets $A$ and $B$, we use $A \times B$ to denote their Cartesian product and use $A-B$ to denote their difference. For any relation $R \subseteq A \times B$ and any $a \in A$, we define $R[a]:=\{b \in B \mid (a, b) \in R\}$.

A (partial) finite state automaton $G$ over alphabet  $\Sigma$ is a 5-tuple $(Q, \Sigma, \delta, q_0, Q_m)$, where $Q$ is the finite set of states, $\delta: Q \times \Sigma \longrightarrow Q$ is the (partial) transition function\footnote{As usual, we also view the partial transition function $\delta: Q \times \Sigma \rightarrow Q$ as a relation $\delta \subseteq Q \times \Sigma \times Q$.}, $q_0 \in Q$ the initial state and $Q_m \subseteq Q$ the set of marked states. We shall write $\delta(q, \sigma)!$ to mean  $\delta(q, \sigma)$ is defined. 
When $Q_m=Q$, we  also write $G=(Q, \Sigma, \delta, q_0)$.  
As usual, for any $G_1=(Q_1, \Sigma_1, \delta_1, q_{1,0}, Q_{1, m}), G_2=(Q_2, \Sigma_2, \delta_2, q_{2,0}, Q_{2, m})$, we  write $G:=G_1 \lVert G_2$ to denote their synchronous product. We have 
$G = (Q := Q_1\times Q_2,\Sigma:=\Sigma_1\cup\Sigma_2,\delta := {\delta}_1 \lVert {\delta}_2,q_0:=(q_{1,0},q_{2,0}), Q_m:=Q_{1, m} \times Q_{2, m})$, 
where the (partial) transition function $\delta$ is defined as follows:
for any $q = (q_1,q_2)\in Q$ and any\footnote{For example, if $\sigma \in \Sigma_1- \Sigma_2$ and $ \delta_1(q_1, \sigma)$ is undefined, we treat $\delta(q, \sigma)$ as undefined. This convention is adopted throughout the work.} $\sigma \in \Sigma$, 
\begin{center}
$ \delta(q,\sigma):=\left\{
\begin{array}{rcl}
({\delta}_1(q_1,\sigma),q_2), && \text{if } {\sigma \in {\Sigma}_1}- {\Sigma}_2 \\
(q_1,{\delta}_2(q_2,\sigma)), && \text{if } {\sigma \in {\Sigma}_2}- {\Sigma}_1 \\
({\delta}_1(q_1,\sigma),{\delta}_2(q_2,\sigma)), && \text{if } {\sigma \in {\Sigma}_1}\cap {\Sigma}_2 \\
\end{array} \right. $
\end{center}
\noindent
For each sub-alphabet $\Sigma' \subseteq \Sigma$, the natural projection $P_{\Sigma'}: \Sigma^* \rightarrow \Sigma'^*$ is defined, which is  extended to a mapping between languages as usual~\cite{WMW10}. Let $G=(Q, \Sigma, \delta, q_0)$. We abuse the notation and define $P_{\Sigma'}(G)$ to be the finite  automaton $(2^Q, \Sigma, \Delta, UR_{G, \Sigma-\Sigma'}(q_0))$, where the unobservable reach $UR_{G, \Sigma-\Sigma'}(q_0):=\{q \in Q \mid  \exists s \in (\Sigma-\Sigma')^*, q=\delta(q_0, s)\} \in 2^Q$ of $q_0$ with respect to the sub-alphabet\footnote{If $\Sigma=\Sigma'$, then we have $UR_{G, \varnothing}(q_0)$, which is by definition equal to $\{q_0\}$.} $\Sigma-\Sigma' \subseteq \Sigma$ is the initial state (of $P_{\Sigma'}(G)$), and the partial transition function $\Delta: 2^Q \times \Sigma \longrightarrow 2^Q$ is defined as follows. 
\begin{enumerate}
\item for any $\varnothing \neq Q' \subseteq Q$ and any $\sigma \in \Sigma'$, $\Delta(Q', \sigma)=UR_{G, \Sigma-\Sigma'}(\delta(Q', \sigma))$, where we define $UR_{G, \Sigma-\Sigma'}(Q''):=\bigcup_{q \in Q''}UR_{G, \Sigma-\Sigma'}(q)$ for any $Q'' \subseteq Q$.
\item for any $\varnothing \neq Q' \subseteq Q$ and any $\sigma \in \Sigma-\Sigma'$, $\Delta(Q' , \sigma)=Q'$. 
\end{enumerate}
We here shall emphasize that $P_{\Sigma'}(G)$ is over  $\Sigma$, instead of $\Sigma'$, and there is no transition defined at the state $\varnothing \in 2^Q$. 

For any finite state automaton $G=(Q, \Sigma, \delta, q_0, Q_m)$, we write $L(G)$ and $L_m(G)$ to denote the closed-behavior and the marked-behavior of $G$~\cite{WMW10}, respectively. 


\section{System Setup, Model Constructions and Synthesis Solution}
\subsection{System Setup}
We shall first introduce and present a formalization of the system components, mostly following~\cite{Lin2018, LZS19, LS20, LS20J}. We adapt the  water tank example of~\cite{Su2018} as a running example to illustrate the constructions and effectiveness of our approach.  

{\bf Plant}: The plant is given by a finite state automaton $G=(Q, \Sigma, \delta, q_0)$. Let $Q_{bad} \subseteq Q$ denote the set of bad states for $G$. Without loss of generality, we shall assume each  state in $Q_{bad}$ is  deadlocked, since damage cannot be undone. Thus, we can merge all the $Q_{bad}$ states into an equivalent   state $q_{bad} \in Q$. In the rest, we shall let $q_{bad}$ denote the unique bad state for $G$. Without loss of generality, we shall assume $q_{bad} \neq q_0$. We let $\Sigma_o \subseteq \Sigma$ denote the subset of observable events and  $\Sigma_c \subseteq \Sigma$  denote the subset of controllable events for the supervisor. We  shall refer to the tuple $(\Sigma_c, \Sigma_o)$  as a control constraint. As usual, let $\Sigma_{uo}=\Sigma-\Sigma_o$ denote the subset of unobservable events and   $\Sigma_{uc}=\Sigma-\Sigma_c$ denote the subset of uncontrollable events. 

\begin{figure}[h]
    \centering
    \includegraphics[scale = 0.55]{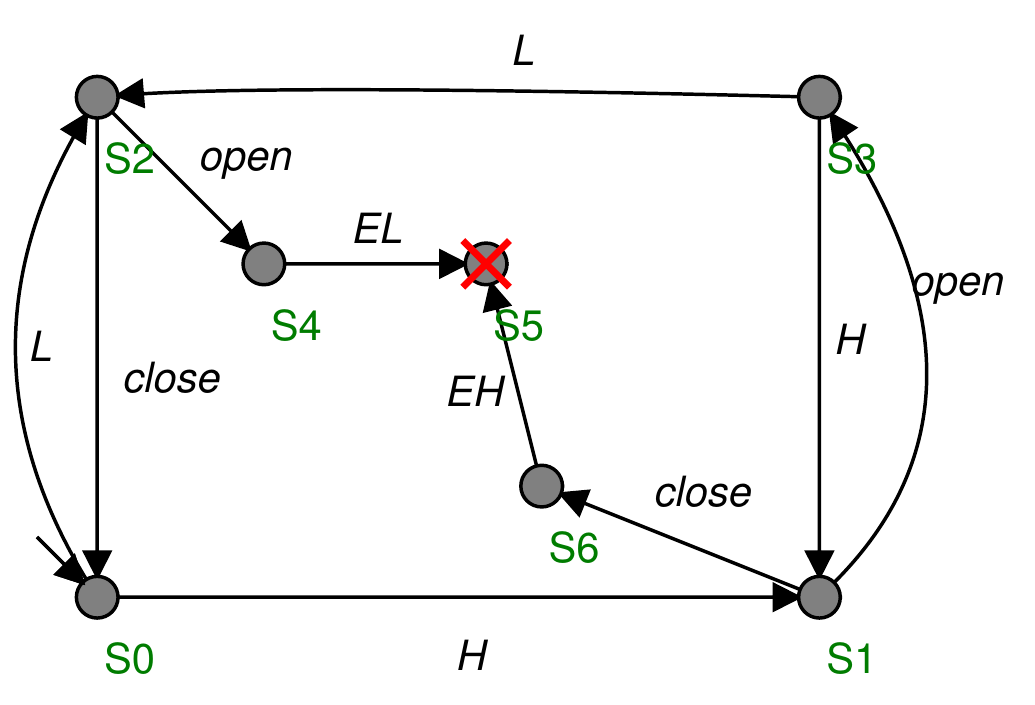}
    \caption{The  plant $G$ with the bad (crossed) state for the water tank example}
    \label{fig:plant}
\end{figure}

\begin{example}
The water tank system~\cite{Su2018} has a constant supply rate, a water tank, and a control valve at the bottom of
the tank  controlling the outgoing flow rate. We assume  the valve can only be fully open or
fully closed. The water level  could be measured, whose
value can trigger some predefined events that denote the
water levels: low ($L$), high ($H$), extremely low ($EL$) and extremely high
($EH$). The model of the plant $G$ is shown in Fig.~\ref{fig:plant} and $S5$ is the  bad state $q_{bad}$ which is crossed. We assume all the events are observable to the supervisor, i.e., $\Sigma_o=\Sigma=\{L, H, EL, EH, open, close\}$; only the events of opening the
valve and  closing the valve are controllable to the supervisor, i.e., $\Sigma_c=\{open, close\}$. 
\end{example}

{\bf Supervisor}:  
 In the absence of an attacker, a supervisor over the control constraint $(\Sigma_c, \Sigma_o)$ is  often modelled as a finite state automaton $S=(X, \Sigma, \zeta, x_0)$, which  satisfies the  controllability and observability constraints~\cite{B1993}:
\begin{enumerate}
\item [$\bullet$] ({\em controllability}) for any state $x \in X$ and any uncontrollable event $\sigma \in \Sigma_{uc}$, $\zeta(x, \sigma)!$,
\item [$\bullet$] ({\em observability}) for any state $x \in X$ and any unobservable event $\sigma \in \Sigma_{uo}$, $\zeta(x, \sigma)!$ implies $\zeta(x, \sigma)=x$.
\end{enumerate}
The control command issued by supervisor $S$ at state $x \in X$ is defined to be $\Gamma(x):=\{\sigma \in \Sigma \mid \zeta(x, \sigma)!\}$. We assume the supervisor $S$ will issue a control command to the plant whenever  an observable event is received and when the supervisor is initiated at the initial state.   Let $\Gamma:=\{\gamma \subseteq \Sigma \mid \Sigma_{uc} \subseteq \gamma\}$ denote the set of all the possible control commands.

\begin{figure}[h]
    \centering
    \includegraphics[scale = 0.5]{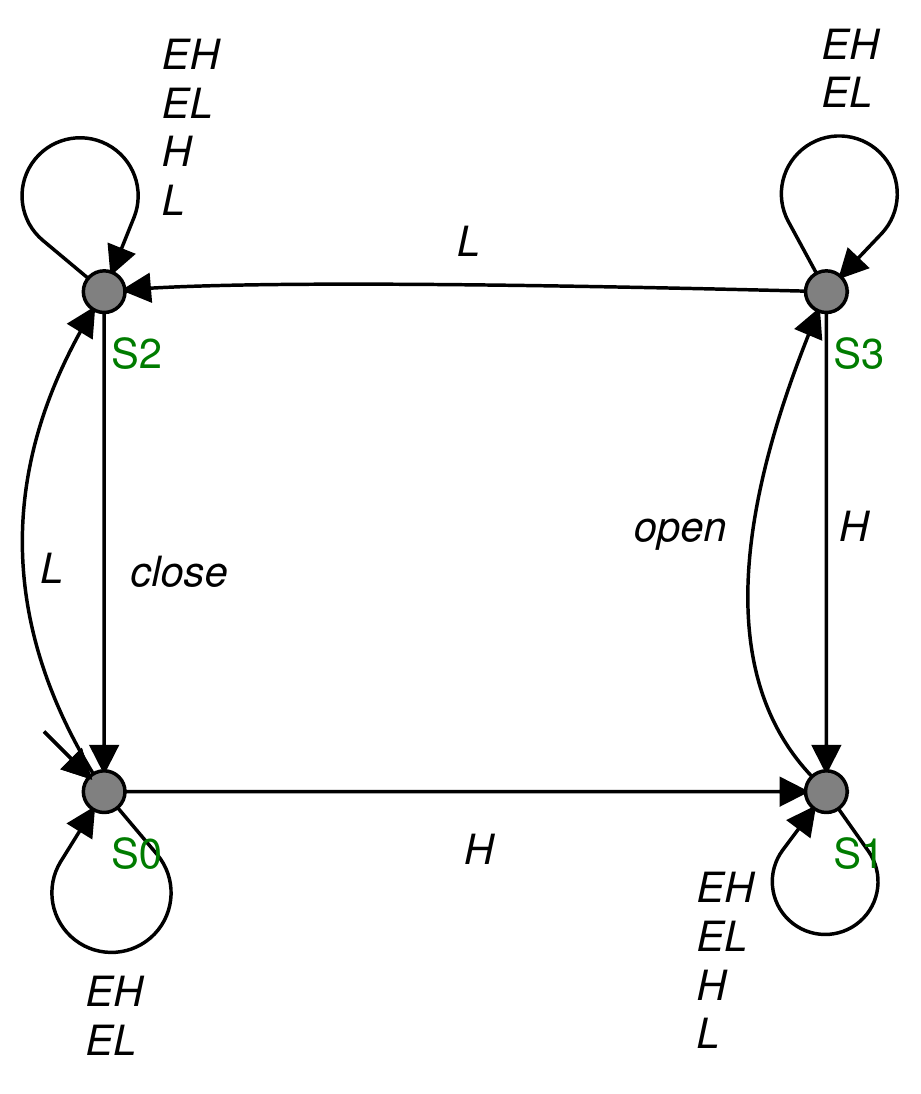}
    \caption{The supervisor $S$ for the water tank example}
    \label{fig:supervisor}
\end{figure}

\begin{example}
We assume a supervisor $S$ has been synthesized to control $G$ in the water tank example. The model of the supervisor $S$ is shown in  Fig.~\ref{fig:supervisor}.  We shall remark that the supervisor $S$  prevents the water level from becoming extremely high (respectively, extremely low), by opening (respectively, closing) the
valve when the water level is high (respectively, low). 
\end{example}

{\bf Observation Automaton}: The adversary has  recorded a (prefix-closed) finite set $O \subseteq P_{\Sigma_o}(L(S \lVert G))$ of observations of the runs of the closed-loop system $S \lVert G$, where $P_{\Sigma_o}: \Sigma^* \rightarrow \Sigma_o^*$ denotes the natural projection~\cite{WMW10}. $O$ is given  by  an automaton $M_O=(U, \Sigma_o, \eta, u_0)$, i.e.,  $O=L(M_O)$. We refer to $M_O$ as an observation automaton. Without loss of generality, we shall assume 
there is exactly one deadlocked state $u_{\bot} \in U$ in $M_O$ and, for any maximal string $s \in O$ (in the prefix ordering~\cite{WMW10}), we have $\eta(u_0, s) =u_{\bot}$. Any supervisor $S'$ that can generate such observations $O$, i.e., $O \subseteq P_{\Sigma_o}(L(S' \lVert G))$, is said to be consistent with $O$.

\begin{figure}[h]
    \centering
    \includegraphics[scale = 0.5]{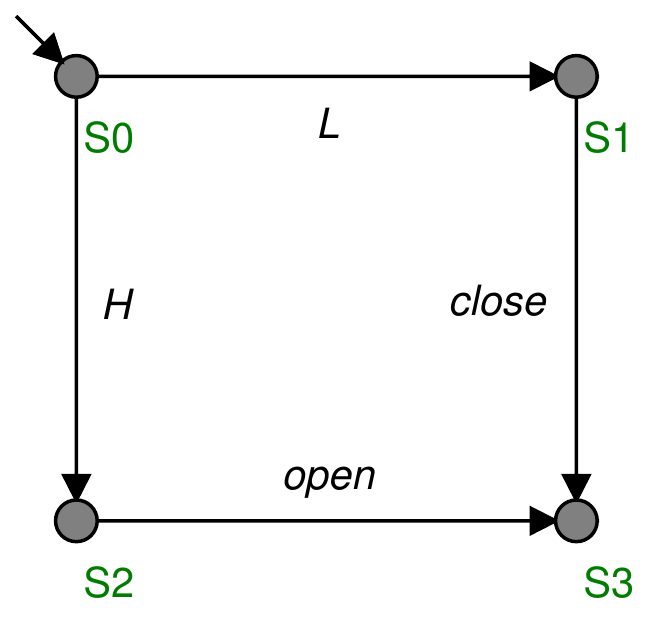}
    \caption{The  observation automaton $M_O$ for the water tank example}
    \label{fig:observation}
\end{figure}

\begin{example}
Let us continue with the water tank example. The observation automaton $M_O$ is given in Fig.~\ref{fig:observation}. It is clear that $O \subseteq P_{\Sigma_o}(L(S\lVert G))$. Thus, $S$ is consistent with $O$. 
\end{example}

{\bf Monitor}:  We assume there exists a monitor $M$ that  records its observation $w \in \Sigma_o^*$ of the execution of the (attacked) closed-loop system and halts the system execution after the detection of an attacker~\cite{LZS19, Lin2018, LS20,  LS20J}. It will conclude the existence of an attacker (at the first moment) when it observes
some string $w \notin P_{\Sigma_o}(L(S \lVert G))$. We  remark that some string $w \notin P_{\Sigma_o}(L(S \lVert G))$ has been generated if and only if $P_{\Sigma_o}(S \lVert G)$ reaches the $\varnothing \in 2^{X \times Q}$ state~\cite{LZS19}.  We here shall refer to $P_{\Sigma_o}(S \lVert G)$ as the monitor.  That is, 
\begin{center}$M=P_{\Sigma_o}(S\lVert G)=(2^{X \times Q}, \Sigma, \Delta, UR_{S\lVert G, \Sigma-\Sigma_o}(x_0, q_0))$.\end{center} 
The monitor state $\varnothing \in 2^{X \times Q}$ and any plant state $q \neq q_{bad}$ together defines the covertness-breaking states for the attacker~\cite{LZS19, LS20, LS20J}.

\begin{figure}[h]
    \centering
    \includegraphics[scale = 0.55]{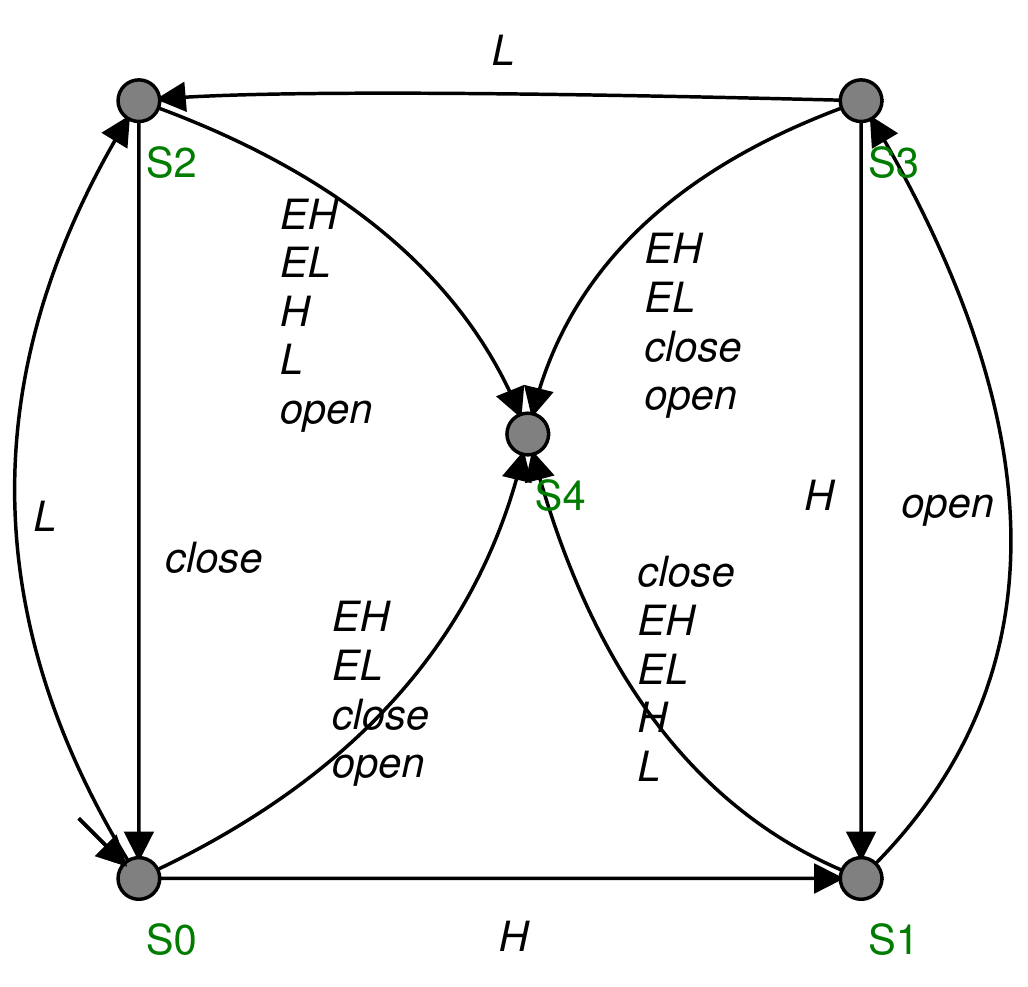}
    \caption{The  monitor $M$ for the water tank example}
    \label{fig:monitor}
\end{figure}

\begin{example}
Let us continue with the water tank example, with the model of the plant $G$ given in Fig. 1 and the model of the supervisor $S$ given in Fig. 2. Then, the monitor $M=P_{\Sigma_o}(S\lVert G)$ is given in Fig.~\ref{fig:monitor}, where $S4$ denotes the state $\varnothing \in 2^{X \times Q}$.
\end{example}

{\bf Attacker}: In this paper, we assume  $\Sigma_o$ is also the subset of plant events that can be observed by the attacker. Let $\Sigma_{a, A} \subseteq \Sigma_c$ denote the subset of controllable events that can be compromised under actuator (disablement) attacks.
Let $\Sigma_{s, A} \subseteq \Sigma_o$ denote the subset of observable events that can be compromised under sensor attacks. We adopt a  relation $R \subseteq \Sigma_{s, A} \times \Sigma_{s, A} $ to specify the sensor  attack capabilities~\cite{LS20}. Intuitively, any  event $\sigma \in \Sigma_{s, A}$ executed in the plant $G$ may lead to the observation of some event $\sigma'$ in $R[\sigma] =\{\sigma' \in \Sigma_{s, A}  \mid (\sigma, \sigma') \in R\}$ by the supervisor $S$, due to the  attacker exercising the sensor replacement attack. Without loss of generality, we shall  assume $\sigma \in R[\sigma]$ and  $R[\sigma]-\{\sigma\} \neq \varnothing$, for any $\sigma \in \Sigma_{s, A}$. 
We shall refer to the tuple $\mathcal{T}=(\Sigma_o, \Sigma_{a, A},     (\Sigma_{s, A}, R))$ as an attack constraint. 

In this paper, we assume the adversary knows the bad state $q_{bad} \in Q$, the model of $G$ and the control constraint $(\Sigma_c, \Sigma_o)$. On the other hand, we assume the model of the supervisor $S$, and thus the model of the monitor $M$, is unknown to the adversary. 

\begin{example}
We assume all the water level events are compromised observable events to the attacker, i.e., $\Sigma_{s, A}=\{L, H, EL, EH\}$, and  $R=\Sigma_{s, A} \times \Sigma_{s, A}$.  We also assume all the controllable events are compromised, i.e., $\Sigma_{a, A}= \Sigma_c=\{open, close\}$. 
\end{example}

\subsection{Model Constructions}
Given the model of the plant $G=(Q, \Sigma, \delta, q_0)$,  with the bad state $q_{bad} \in Q$, the  observation automaton $M_O=(U, \Sigma_o, \eta, u_0)$ and the attack constraint $\mathcal{T}=(\Sigma_o, \Sigma_{a, A},   (\Sigma_{s, A}, R))$, in this paper we will construct the following four automata to perform the synthesis. 
\begin{enumerate}
    \item The transformed plant $G^T$, which reflects a) the bad state $q_{bad}$ of $G$ is the goal state for the attacker, b) before the damage is inflicted, i.e., before  $q_{bad}$ is reached, any uncertainty that might cause the covertness  to be broken is considered to be bad for the attacker, and c) nothing can be executed  after the damage is inflicted.
    \item The unconstrained sensor attack automaton $G_{SA}$, which specifies all the possible   sensor replacement attacks that can be carried out, i.e., upon the receiving of some compromised observable event in $\Sigma_{s, A}$ from the plant, the attacker can issue some attacked copy in $\Sigma_{s, A}^{\#}$, as specified by $R$, to mislead the supervisor. 
    \item The attack-forcing   automaton $G_{AF}$ that forces sensor replacement attacks, which ensures a) some sensor replacement attacks must be carried out in order to fulfill the damage-inflicting goal of the attacker, and b) covertness could be broken once some sensor replacement attacks have been performed\footnote{We here remark  that actuator disablement attacks cannot cause the covertness to be broken, following~\cite{LZS19}. Indeed, the supervisor is not sure whether some event $\sigma \in \Sigma_{c, A}$ has been disabled by an attacker, even if disabling $\sigma$ may result
in deadlock, as the supervisor is never sure whether: 1) deadlock has occurred due to actuator attack, or 2)
$\sigma$ will possibly fire soon (according to the internal mechanism of the plant), without an explicit timing mechanism.}. 
    \item The transformed observation automaton $M_O^T$, which reflects a) any (attacked) observation for the supervisor that falls within $O$ is not bad for the attacker, b) any (attacked) observation for the supervisor that falls outside $O$ is (considered to be) bad for the attacker, if the attacker has already carried out some sensor replacement attacks and  the damage has not been inflicted, and c) the supervisor and the monitor  receive  attacked copies in $\Sigma_{s, A}^{\#}$ for events in $\Sigma_{s, A}$ executed by the plant. 
\end{enumerate}
The idea of the constructions is explained as follows.
\begin{enumerate}
    \item Since  the models of the supervisor $S$  and the monitor $M$ are not available, we do not have the model of  the attacked closed-loop system~\cite{LS20, LS20J}. Thus, we cannot use the  set of  covertness-breaking states of the attacked closed-loop system to perform the synthesis. To ensure the covertness of the synthesized attacker, the idea is to  over-approximate the set of covertness-breaking states of the attacked closed-loop system, without using the models of $S$ and $M$. The over-approximation needs to work for any $S$ (and thus $M$) that is consistent with $O$. In this work, we use 
    \begin{enumerate}
        \item the state $q\neq q_{bad}$ of $G^T$ where damage has not been inflicted,
    \item the state of $G_{AF}$ where sensor replacement attacks have been carried out, and \item the state of $M_O^T$ where an (attacked) observation that falls outside $O$ has been observed by the supervisor (and the monitor)
    \end{enumerate}
    to (together) over-approximate the set of covertness-breaking states. Indeed, if the monitor $M$  reaches the $\varnothing \in 2^{X \times Q}$ state (under attacks), then Conditions b) and c
    ) above must be both satisfied. 
    \item We view  $G^T, M_O^T, G_{SA}, G_{AF}$ as the components of the surrogate plant $G^T \lVert M_O^T \lVert G_{SA}\lVert$ \\ $G_{AF}$ and view the attacker $A$ as a supervisor that controls the surrogate plant to i) avoid breaking the covertness, over-approximated with Conditions a), b) and c), and ii) ensure the damage-infliction, i.e., the reachability of the $q_{bad}$ state. 
    \item  By construction, $G^T \lVert M_O^T \lVert G_{SA}\lVert G_{AF} \lVert A$ can be viewed as providing a (behavioral) upper bound for the attacked closed-loop system of~\cite{LS20J, LS20}. If $G^T \lVert M_O^T \lVert G_{SA}\lVert G_{AF}$ can be controlled to avoid reaching a set that contains all the  covertness-breaking states by the attacker $A$,  then the attacked closed loop system (induced by $A$) can also avoid reaching the  covertness-breaking states. 
    \item The reachability of the  state $q_{bad}$  in $G^T \lVert M_O^T \lVert G_{SA}\lVert G_{AF} \lVert A$ in general does not imply the reachability of the state $q_{bad}$ in the attacked closed-loop system. In particular, by construction, the $q_{bad}$ state in $G$  can already be reached  in  $G^T \lVert M_O^T \lVert G_{SA}\lVert G_{AF}$, since we ignore the control effect of the supervisor that ensures the non-reachability of $q_{bad}$. That is, only some of the executions that lead to the  state $q_{bad}$ in $G^T \lVert M_O^T \lVert$ \\$ G_{SA}\lVert G_{AF} \lVert A$  indeed exist in the attacked closed-loop system. In order to address this issue,  we carefully design the state markings for the new plant $G^T \lVert M_O^T \lVert G_{SA}\lVert $\\$ G_{AF}$  to ensure that a marked state is reached in $G^T \lVert M_O^T \lVert G_{SA}\lVert G_{AF}$ if and only if i) the bad state $q_{bad} \in Q$ has been reached, ii) some sensor replacement attacks have been carried out (before the damage is inflicted). It turns out that the synthesized attacker can ensure the damage-reachability in the attacked closed-loop system if there exists a marked string $s \in L_m(G^T \lVert M_O^T \lVert G_{SA}\lVert G_{AF} \lVert A)$ such that $s$ is allowed by an attacked supervisor $S^{\downarrow, A}$ where $S^{\downarrow}$ under-approximates any supervisor that is consistent with $O$. 
    It follows that $S^{\downarrow, A}$ can be used in the verification of damage-reachability or even in the synthesis for ensuring damage-reachability by construction.
\end{enumerate}
We are now ready to present the model constructions.

{\bf Transformed Plant}: We model the transformed plant as
\begin{center}
    $G^T=(Q \cup \{q^{\$}\}, \Sigma \cup \Sigma_{s, A}^{\#}\cup \{\$\} \cup \Gamma, \delta^T, q_0, \{q_{bad}\})$,
\end{center}
where $q^{\$} \notin Q$ is a newly added state,  $\Sigma_{s, A}^{\#}=\{\sigma^{\#} \mid \sigma \in \Sigma_{s, A}\}$ is a  relabelled copy of $\Sigma_{s, A}$, $\$ \notin \Sigma \cup \Sigma_{s, A}^{\#} \cup \Gamma$ is a newly added event and $\delta^T: (Q \cup \{q^{\$}\}) \times (\Sigma \cup \Sigma_{s, A}^{\#}\cup \{\$\} \cup \Gamma) \rightarrow (Q \cup \{q^{\$}\})$ is the partial transition function defined as follows.
\begin{enumerate}
\item for any $q \in Q$ and any $\sigma \in \Sigma$, $\delta^T(q, \sigma)=\delta(q, \sigma)$,
\item for any $q \in Q-\{q_{bad}\}$, $\delta(q, \$)=q^{\$}$,
\item for any $q \in Q-\{q_{bad}\}$ and any $\sigma \in \Sigma_{s, A}^{\#} \cup \Gamma$, $\delta(q, \sigma)=q$.
\end{enumerate}
Intuitively, the 
event $\$$ is used to denote that the following conditions hold simultaneously: a) damage has not been inflicted in the plant $G$, b) some sensor replacement attack has been performed, c) the (attacked) observation for the supervisor has fallen outside $O$. In particular, b) and c) implies that the presence of the attacker could have been discovered by the monitor; a), b) and c) together implies that the covertness of the attacker could have been broken. Thus, $\$$ is an uncontrollable (and unobservable) ``bad" event for the attacker that leads to the bad state\footnote{Recall that $q_{bad}$ is the goal state for the attacker.} $q^{\$}$ in $G^T$. Rule 2) here contributes to Condition a) of the definition of $\$$. Intuitively, the supervisor only receives the relabelled copies in $\Sigma_{s, A}^{\#}$,  while the events in $ \Sigma_{s, A}$ are executed in the plant. Rule 3) is added to  ensure that nothing can be executed when the state $q_{bad}$ is reached.
The state size of $G^T$ is $|Q|+1$. 

\begin{figure}[h]
    \centering
    \includegraphics[scale = 0.45]{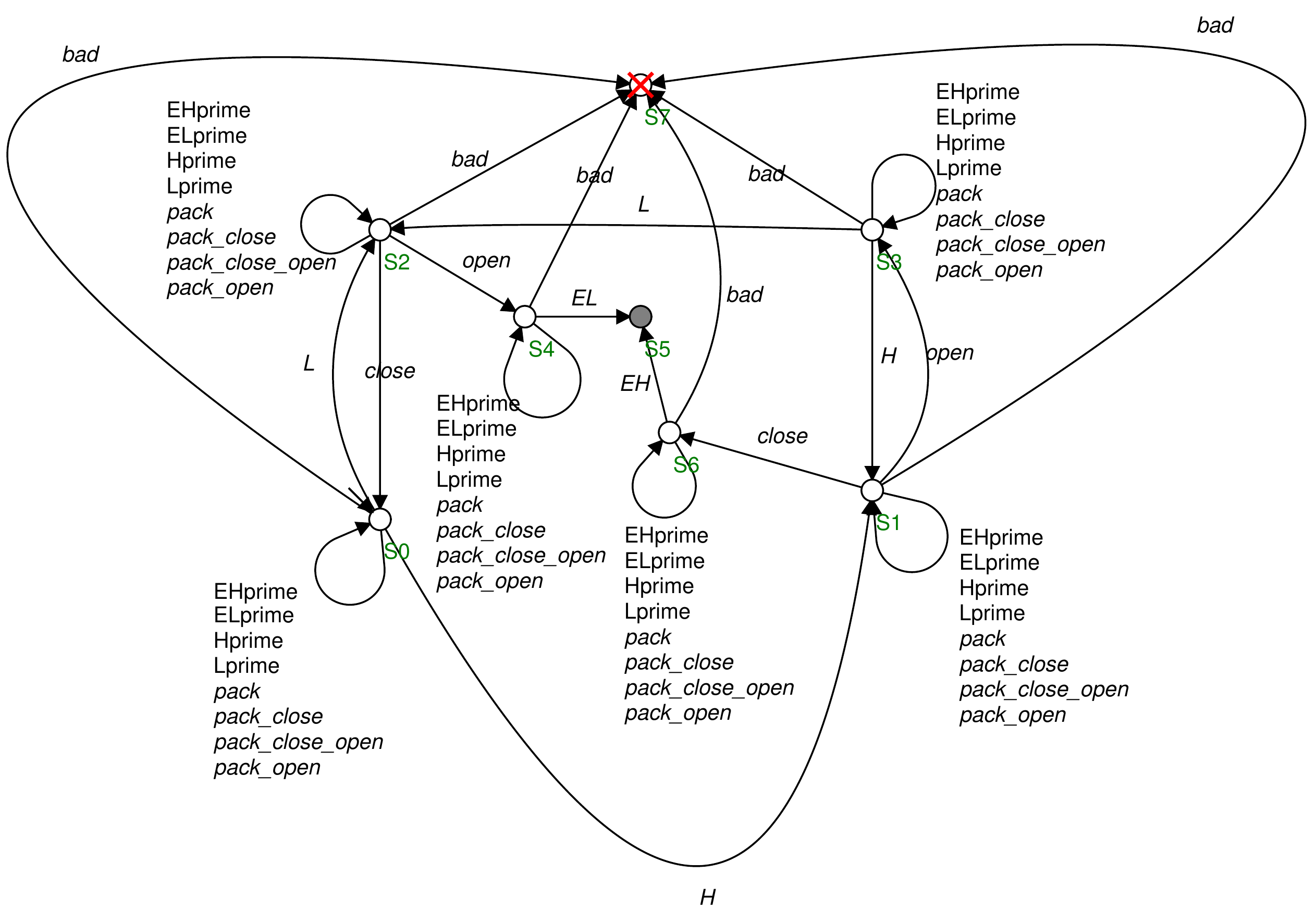}
    \caption{The transformed plant $G^T$}
    \label{fig:tranplant}
\end{figure}

\begin{example}
For the water tank system, the transformed plant $G^T$ is provided in Fig.~\ref{fig:tranplant}, where the event ``bad" is used to represent $\$$; the event $pack$ (respectively, $pack\_open$, $pack\_close$, $pack\_close\_open$) denotes (the receiving of) the control command $\gamma=\Sigma_{uc}$ (respectively, $\gamma= \Sigma_{uc} \cup \{open\}$, $\gamma= \Sigma_{uc} \cup \{close\}$, $\gamma= \Sigma_{uc} \cup \{close, open\}$);  the event $Lprime$ (respectively, $Hprime, ELprime, EHprime$) is then used to denote  $L^{\#}$ (respectively, $H^{\#}, EL^{\#}, EH^{\#}$). In Fig. 5, $S7$ denotes the state $q^{\$}$ and  $S5$ is used to denote the state $q_{bad}$.
\end{example}

{\bf Unconstrained Sensor Attack Automaton:} We  model  the  sensor (replacement) attack capabilities by using a finite state automaton~\cite{LS20} 
\begin{center}$G_{SA}=(Q^{SA},   \Sigma  \cup \Sigma_{s, A}^{\#}, \delta^{SA}, q_{init})$,\end{center} 
where  $Q^{SA}=\{q^{\sigma} \mid \sigma \in \Sigma_{s, A}\} \cup \{q_{init}\}$.   $\delta^{SA}: Q^{SA} \times ( \Sigma  \cup \Sigma_{s, A}^{\#}) \longrightarrow Q^{SA}$ is the partial transition function defined in the following. 
\begin{enumerate}
    \item [1.] for any $\sigma \in \Sigma_{s, A}$, $\delta^{SA}(q_{init}, \sigma)=q^{\sigma}$,
    \item [2.]  for any $q^{\sigma} \in Q^{SA}-\{q_{init}\}$ and for any $\sigma' \in R[\sigma]$, $\delta^{SA}(q^{\sigma}, \sigma'^{\#})=q_{init}$,
    \item [3.]  for any  $\sigma \in \Sigma-\Sigma_{s, A}$, $\delta^{SA}(q_{init}, \sigma)=q_{init}$.
\end{enumerate}
 $G_{SA}$ specifies all the possible attacked copies in $\Sigma_{s, A}^{\#}$ which could be received by the supervisor, due to the sensor replacement attacks, for each compromised observable event $\sigma \in \Sigma_{s, A}$ executed in the plant. The state $q^{\sigma}$, where $\sigma \in \Sigma_{s, A}$, is used to denote that the attacker has just received the compromised observable event $\sigma$, with Rule 1). Rule 2) then forces the attacker to (immediately) make a sensor attack decision, upon receiving each compromised observable event. Rule 3) is added such that $G_{SA}$ is over $\Sigma_{s, A}^{\#} \cup \Sigma$ and no attack could be performed when $\sigma \in \Sigma- \Sigma_{s, A}$ is executed. The state size of $G_{SA}$ is $|\Sigma_{s, A}|+1$, before automaton minimization.

\begin{figure}[h]
    \centering
    \includegraphics[scale = 0.55 ]{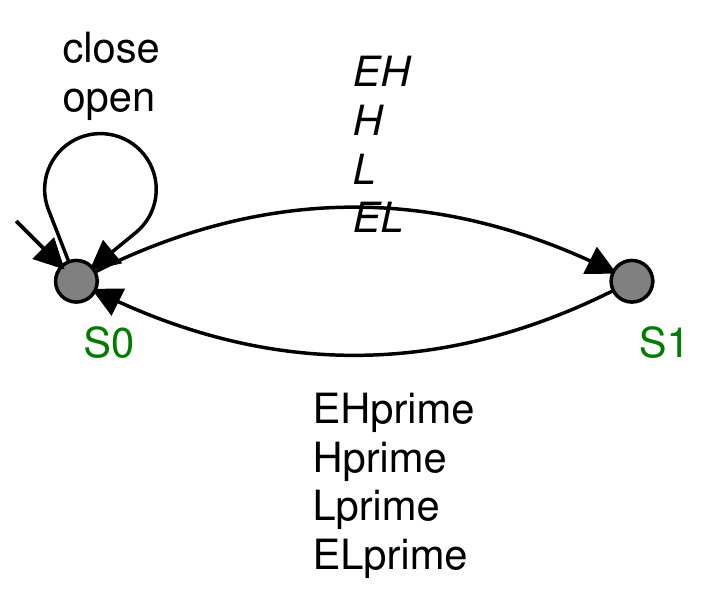}
    \caption{The unconstrained sensor attack automaton $G_{SA}$}
    \label{fig:SA}
\end{figure}

\begin{example}
We now continue with the water tank example. The unconstrained sensor attack automaton $G_{SA}$ is shown in Fig.~\ref{fig:SA}. We note that the automaton $G_{SA}$ has been minimized and all the $q^{\sigma}$ states, where $\sigma \in \Sigma_{s, A}$, have been merged into an equivalent state, i.e., S1. 
\end{example}

{\bf Transformed Observation Automaton}: We model the transformed observation automaton as
\begin{center}
    $M_O^T=(U \cup \{u^{!}, u^{\$}\}, (\Sigma_o-\Sigma_{s, A}) \cup \Sigma_{s, A}^{\#} \cup \{\$\},  \eta^T, u_0, U \cup \{u!\})$,
\end{center}
where $u^!, u^{\$} \notin U$ and $\eta^T: (U \cup \{u^{!}, u^{\$}\}) \times ((\Sigma_o-\Sigma_{s, A}) \cup \Sigma_{s, A}^{\#} \cup \{\$\}) \rightarrow (U \cup \{u^{!}, u^{\$}\})$ is the partial transition function defined as follows.
\begin{enumerate}
    \item for each $u \in U$ and each $\sigma \in \Sigma_{o}-\Sigma_{s, A}$, if $\eta(u, \sigma)!$, then $\eta^T(u, \sigma)=\eta(u, \sigma)$,
    \item for each $u \in U$ and each $\sigma \in \Sigma_{o}-\Sigma_{s, A}$, if $\neg \eta(u, \sigma)!$, then $\eta^T(u, \sigma)=u^!$,
    \item for each $u \in U$ and each $\sigma \in \Sigma_{s, A}$, if $\eta(u, \sigma)!$, then $\eta^T(u, \sigma^{\#})=\eta(u, \sigma)$,
    \item for each $u \in U$ and each $\sigma \in \Sigma_{s, A}$, if $\neg \eta(u, \sigma)!$, then $\eta^T(u, \sigma^{\#})=u^!$,
    \item for each $\sigma \in \Sigma_o-\Sigma_{s, A}$, $\eta^T(u^!, \sigma)=u^{!}$,
    \item for each $\sigma \in \Sigma_{s, A} $, $\eta^T(u^!, \sigma^{\#})=u^{!}$,
    \item $\eta^T(u^!, \$)=u^{\$}$.
\end{enumerate}
Intuitively, the state $u^!$ here means that the (attacked) observation of the supervisor has fallen outside $O$ and the attacker may be in the risk of exposing itself.  Upon the execution of the ``bad" event $\$$ for the attacker, the bad state $u^{\$}$ of $M_O^T$ can be reached from $u^!$. In particular, Rule 7)  contributes to Condition c) of the definition of $\$$. The state size of $M_O^T$ is $|U|+2$. We remark that $L_m(M_O^T)=((\Sigma_o-\Sigma_{s, A}) \cup \Sigma_{s, A}^{\#})^*$.

\begin{figure}[h]
    \centering
    \includegraphics[scale = 0.5]{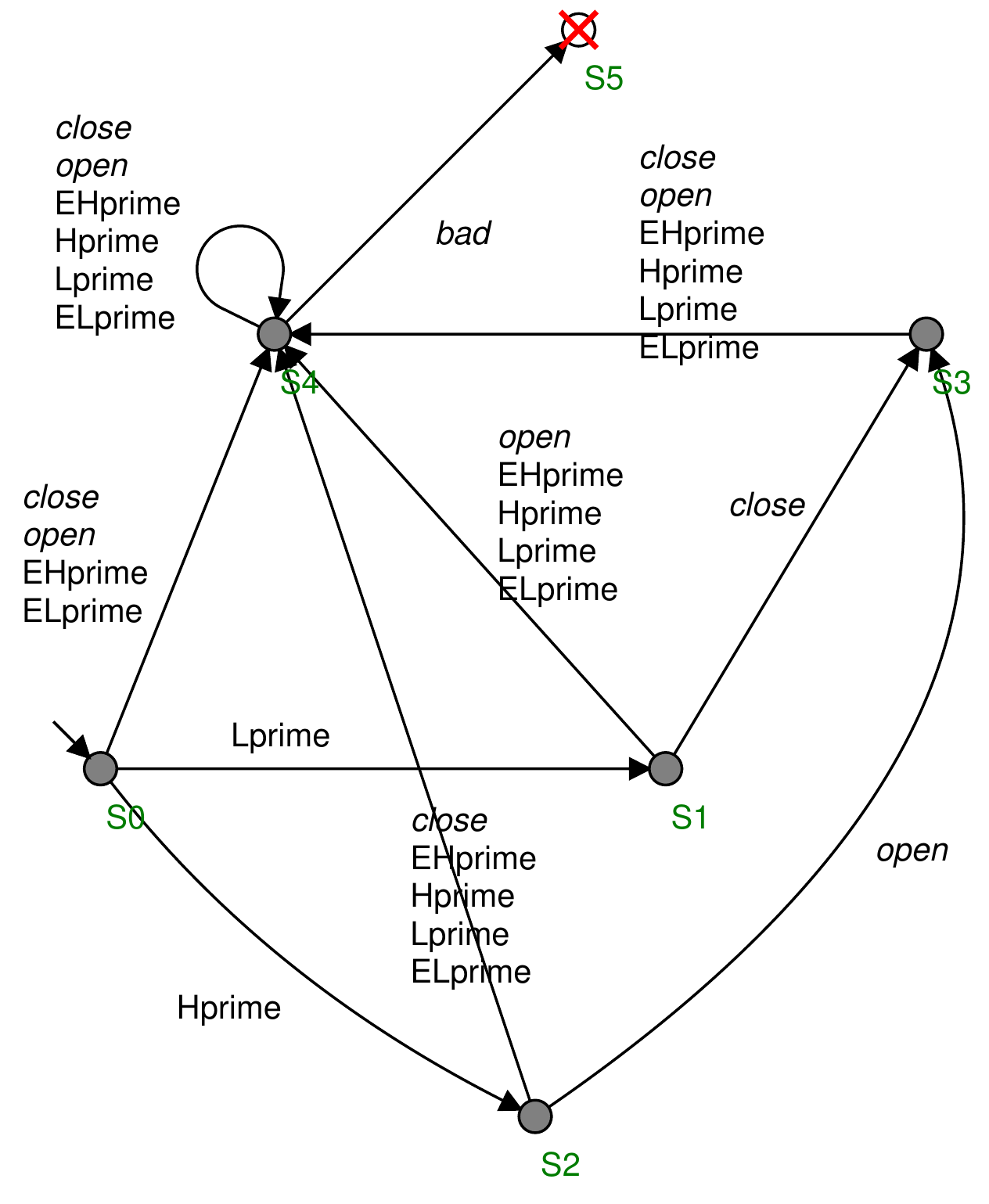}
    \caption{The transformed observation automaton $M_O^T$}
    \label{fig:tranobservation}
\end{figure}

\begin{example}
We now continue with the water tank example. The transformed observation automaton $M_O^T$ is given in Fig.~\ref{fig:tranobservation}. $S4$ denotes the state $u^!$ and $S5$ denotes the state $u^{\$}$.
\end{example}

{\bf Attack-Forcing Automaton: } We model the attack-forcing automaton as 
\begin{center}
    $G_{AF}= (Q^{AF}, \Sigma_{s, A} \cup \Sigma_{s, A}^{\#} \cup \{\$\}, \delta^{AF}, q_0^{AF}, \{q^{AF,!}\})$,
\end{center}
where $Q^{AF}=\{q^{AF, \sigma} \mid \sigma \in \Sigma_{s, A}\} \cup \{q_0^{AF}, q^{AF,!}, q^{AF, \$}\}$ and $\delta^{AF}: Q^{AF} \times (\Sigma_{s, A} \cup \Sigma_{s, A}^{\#} \cup \{\$\}) \rightarrow Q^{AF}$ is the partial transition function defined as follows.
\begin{enumerate}
    \item for any $\sigma \in \Sigma_{s, A}$, $\delta^{AF}(q_0^{AF}, \sigma)=q^{AF,\sigma}$,
    \item for any $\sigma \in \Sigma_{s, A}$, $\delta^{AF}(q^{AF, \sigma}, \sigma^{\#})=q_0^{AF}$,
    \item for any $\sigma, \sigma' \in \Sigma_{s, A}$ with $\sigma \neq \sigma'$, $\delta^{AF}(q^{AF, \sigma}, \sigma'^{\#})=q^{AF, !}$,
    \item $\delta^{AF}(q^{AF,!}, \$)=q^{AF, \$}$.
\end{enumerate}
The state $q^{AF, \sigma}$, where $\sigma \in \Sigma_{s, A}$, is used to denote that the attacker just receives the compromised observable event $\sigma$, with Rule 1). Rule 2) states that the attacker does not effectively carry out sensor replacement attacks and thus returns to the initial state $q_0^{AF}$. Rule 3) captures the situation that the attacker has effectively carried out some sensor replacement attacks. Rule 4) says that, upon the execution of the ``bad" event $\$$ for the attacker, the bad state $q^{AF, \$}$ in $G_{AF}$ can be reached from $q^{AF, !}$.
In particular, Rule 4) here contributes to Condition b) of the definition of $\$$. The state size of $G_{AF}$ is $|\Sigma_{s, A}|+3$.

\begin{figure}[h]
    \centering
    \includegraphics[scale = 0.45]{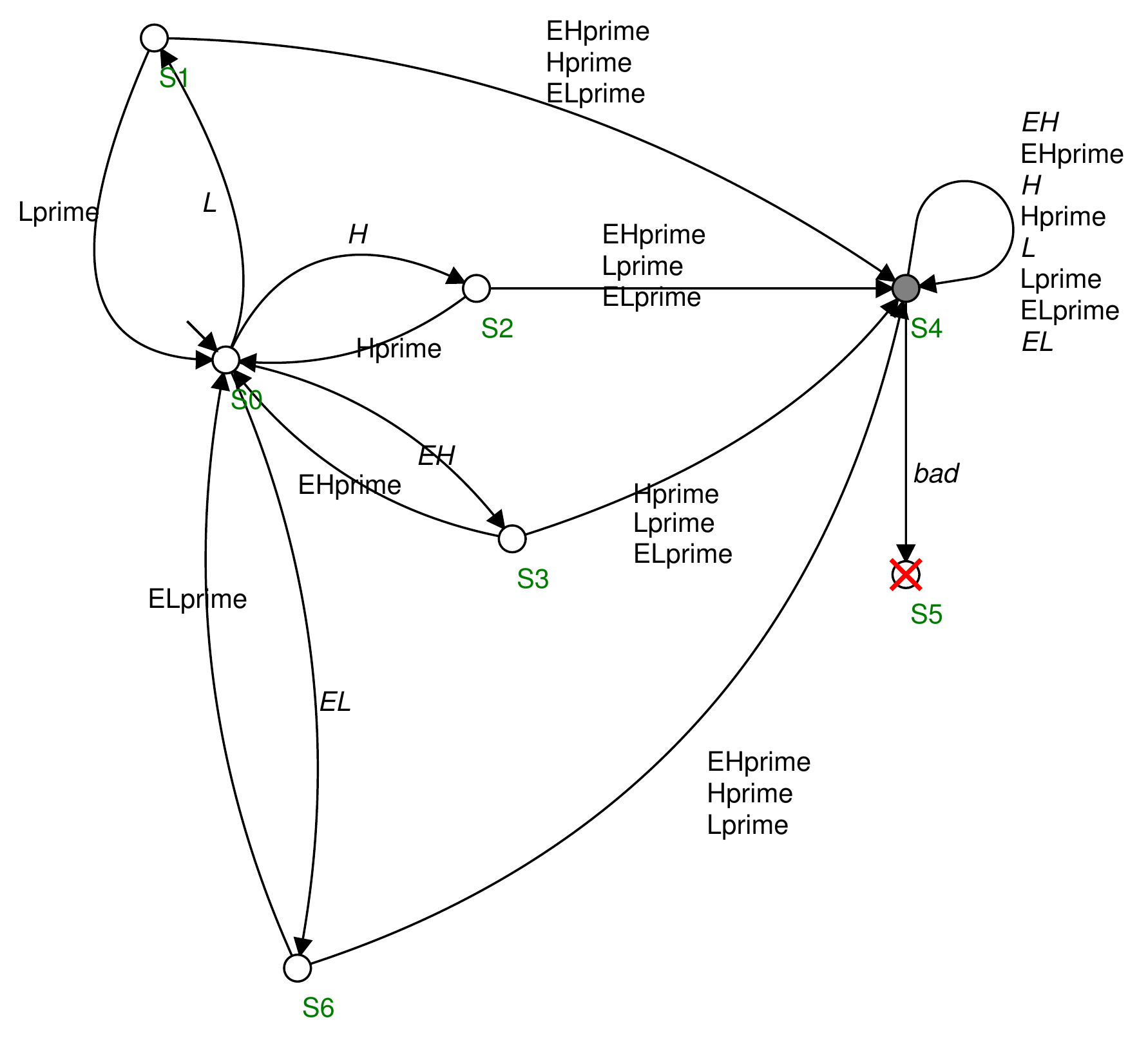}
    \caption{The attack-forcing automaton $G_{AF}$}
    \label{fig:attack}
\end{figure}

\begin{example}
We continue with the water tank example.  The attack-forcing automaton $G_{AF}$ is given in Fig.~\ref{fig:attack}. Here, $S4$ denotes $q^{AF, !}$ and $S5$ denotes $q^{AF, \$}$.
\end{example}

{\bf Attacker}: 
An attacker over  $\mathcal{T}=(\Sigma_o, \Sigma_{a, A},   (\Sigma_{s, A}, R))$ is modelled by a supervisor $A=(Y,  \Sigma \cup \Sigma_{s, A}^{\#}  \cup \{\$\} \cup \Gamma, \beta, y_0)$ over the control constraint $(\Sigma_{a, A} \cup \Sigma_{s, A}^{\#}, \Sigma_o \cup \Sigma_{s, A}^{\#})$. 
Intuitively, the attacker $A$ can only control events in $\Sigma_{a, A} \cup \Sigma_{s, A}^{\#}$ and can only observe events in $\Sigma_o \cup  \Sigma_{s, A}^{\#}$. In this paper, we assume the attacker cannot observe the control commands issued by the supervisor.
\subsection{Synthesis Solution}
\label{sec: SS}
\begin{figure}[h]
    \centering
    \includegraphics[scale = 0.3]{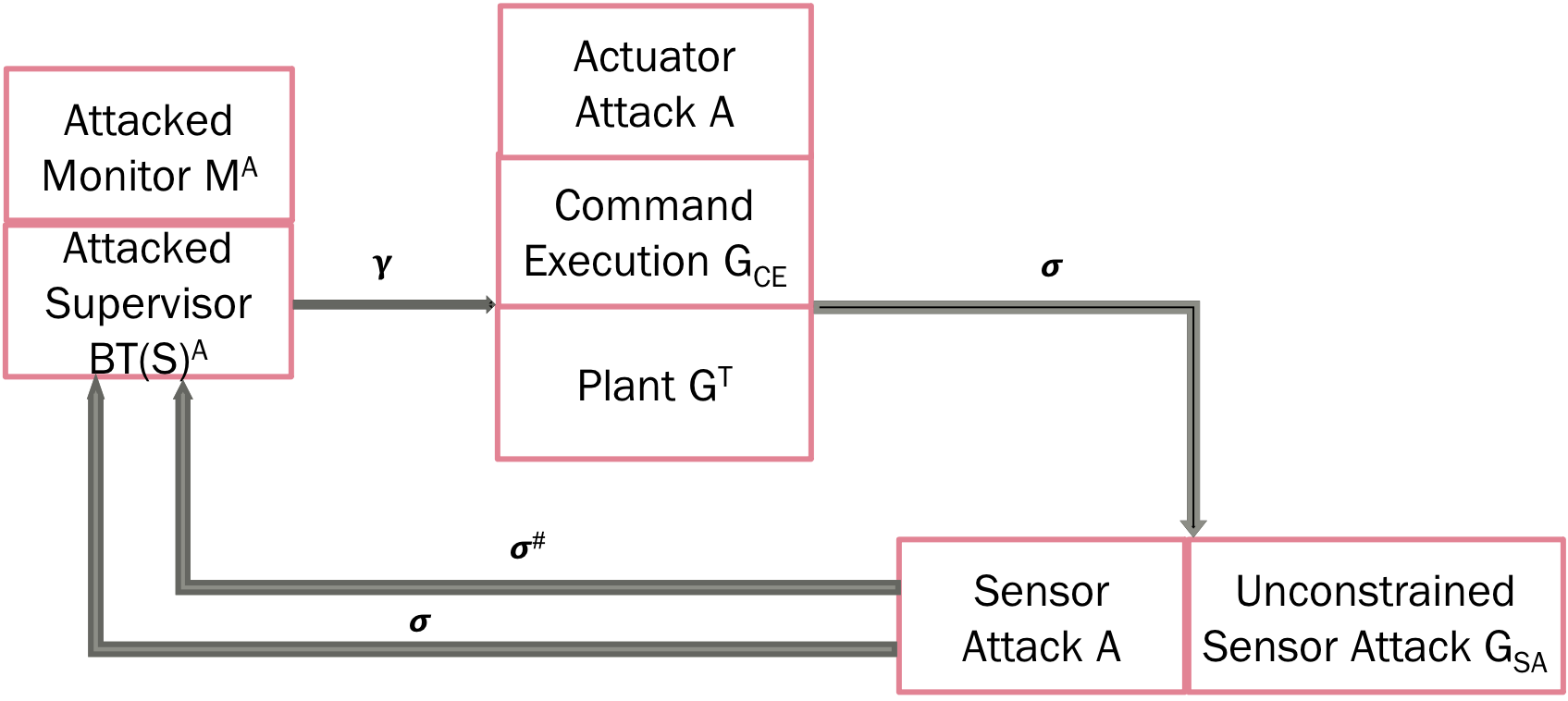}
    \caption{The diagram of the attacked closed-loop system}
    \label{fig:aclp}
\end{figure}

As we have discussed before, we could view  $G^T, M_O^T, G_{SA}, G_{AF}$ as the components of the surrogate plant $G^T \lVert M_O^T \lVert G_{SA}\lVert G_{AF}$ and  view the attacker $A$ as a  supervisor over the control constraint $(\Sigma_{a, A} \cup \Sigma_{s, A}^{\#}, \Sigma_{o}  \cup \Sigma_{s, A}^{\#})$ which controls the surrogate plant. We can then synthesize (maximally permissive) safe attackers for $G^T \lVert M_O^T \lVert G_{SA}\lVert G_{AF}$ by using existing synthesis tools~\cite{Susyna},~\cite{Feng06},~\cite{Malik07}, where  ${\bf BAD}:=\{(q^{\$}, u^{\$}, q^{SA}, q^{AF,\$})\mid q^{SA} \in Q^{SA}\}$ denotes the set of bad states for the attacker $A$ in $G^T \lVert M_O^T \lVert G_{SA}\lVert G_{AF}$. Let $A^{o}$
denote any non-empty (maximally permissive) safe attacker, if it exists, synthesized by using any existing synthesis tool. We still need to ensure the correctness of  $A^o$ for the attacked closed-loop system~\cite{LS20}. To that end, we need to introduce the model of the attacked closed-loop system, induced by $A^o$, which is   $G^T\lVert BT(S)^{A} \lVert M^A \lVert G_{SA} \lVert G_{CE}  \lVert A^o$ (see Fig.~\ref{fig:aclp}), where $BT(S)^{A}$ is the attacked supervisor (with an explicit control command sending phase), $M^A=P_{\Sigma_o}(S \lVert G)^A$ is the attacked monitor and $G_{CE}$ is the command execution automaton (which transduces control command $\gamma$ from the supervisor into event $\sigma \in \Sigma$ executed in the plant). Thus, we still need to introduce the three components $BT(S)^{A}$, $G_{CE}$ and $M^A$ from~\cite{LS20},~\cite{LS20BJ}.

{\bf Attacked Supervisor}: The attacked supervisor $BT(S)^{A}$  is constructed from $S$ as follows. Let
\begin{center}
    $BT(S)^{A}=(X \cup X_{com} \cup \{x^!\}, \Sigma \cup \Sigma_{s, A}^{\#} \cup  \Gamma, \zeta^{BT, A}, x_{0,com})$,
\end{center}
where $X_{com}=\{x_{com} \mid x \in X\}$ is a relabelled copy of $X$, with $X \cap X_{com}=\varnothing$, $x_{0, com} \in X_{com}$ is the relabelled copy of $x_0 \in X$ and  $x^{!} \notin X \cup X_{com}$ is a  newly added state.  The partial transition function $\zeta^{BT, A}$ is defined as follows. 
\begin{enumerate}
    \item  for any $x \in X$,  $\zeta^{BT, A}(x_{com},\Gamma(x))=x$,
    \item for any $x \in X$ and any $\sigma \in \Sigma_{uo}$, if $\zeta(x,\sigma)!$, then  $\zeta^{BT, A}(x, \sigma)=\zeta(x,\sigma)=x$,
    \item for any $x \in X$ and any $\sigma \in \Sigma_{o}-\Sigma_{s, A}$,  if $\zeta(x,\sigma)!$, then $\zeta^{BT, A}(x, \sigma)=\zeta(x,\sigma)_{com}$, with $\zeta(x,\sigma)_{com}$ denoting the  relabelled copy of $\zeta(x,\sigma)$,
    \item for any $x \in X$ and any $\sigma \in \Sigma_{s, A}$, if $\zeta(x,\sigma)!$, then $\zeta^{BT, A}(x, \sigma^{\#})=\zeta(x,\sigma)_{com}$,
    \item  for any $x \in X$ and any $\sigma \in \Sigma_{s, A}$, if $\neg \zeta(x, \sigma)!$, then $\zeta^{BT,A}(x, \sigma^{\#})=x^!$,
    \item for any $x \in X \cup X_{com}$ and any $\sigma \in \Sigma_{s, A}$, $\zeta^{BT,A}(x, \sigma)=x$.
\end{enumerate}
Intuitively, each $x_{com}$ is the control state  corresponding to $x$, which is ready to issue the control command $\Gamma(x)$. In Rule 1), the transition $\zeta^{BT, A}(x_{com},\Gamma(x))=x$ represents the event that the supervisor sends the control command $\Gamma(x)$ to the plant. Each $x \in X$ is a reaction state which is ready to react to an event $\sigma \in \Gamma(x)$  executed by the plant $G$. For any $x \in X$ and any $\sigma \in \Sigma$, the supervisor reacts to the corresponding $\sigma$ transition (fired in the plant) if $\zeta(x, \sigma)!$: if $\sigma \in \Sigma_{uo}$, then the supervisor observes nothing and it remains in the same reaction state $\zeta(x,\sigma)=x \in X$, as defined in Rule 2); if $\sigma \in \Sigma_o$, then the supervisor proceeds to the next control state $\zeta(x,\sigma)_{com}$ and is ready to issue a new control command. Since the supervisor reacts to those events in $\Sigma_{s, A}^{\#}$, instead of the events in $\Sigma_{s, A}$, we need to divide the case $\sigma \in \Sigma_o$ into the case $\sigma \in \Sigma_o-\Sigma_{s, A}$ and the case $\sigma \in \Sigma_{s, A}$ with Rule 3) and Rule 4), respectively. Thus, Rules 1)-4) together captures the control logic of the supervisor $S$ and makes the control command sending phase explicit. 
Rule 5) specifies the situation when the sensor replacement attacks can lead to the state $x^!$, where the existence of attacker is detected (based on the structure of the supervisor alone). Rule 6) is added here such that $BT(S)^{A}$ is over $\Sigma \cup \Sigma_{s, A}^{\#} \cup  \Gamma$  and no event in $\Sigma \cup \Sigma_{s, A}^{\#} \cup  \Gamma$ can be executed at state $x^!$, where the system execution is halted. 

{\bf Attacked Monitor:} 
The attacked monitor $M^A$ is constructed from the monitor $M$ as follows. Let \begin{center}$M^A=P_{\Sigma_o}(S\lVert G)^A=(2^{X \times Q} \cup \{D^{\$}\}, \Sigma \cup \Sigma_{s, A}^{\#} \cup  \{\$\} \cup \Gamma, \Delta^A, UR_{S\lVert G, \Sigma-\Sigma_o}(x_0, q_0), 2^{X \times Q}),$\end{center}
where $D^{\$} \notin 2^{X \times Q}$ is the distinguished bad state of $M^A$ and $\Delta^A: (2^{X \times Q} \cup \{D^{\$}\}) \times (\Sigma \cup \Sigma_{s, A}^{\#} \cup  \{\$\} \cup \Gamma) \rightarrow (2^{X \times Q} \cup \{D^{\$}\})$ is the partial transition function defined as follows.
\begin{enumerate}
\item for any $D \subseteq X \times Q$ and for any $\sigma \in \Sigma-\Sigma_{s, A}$, $\Delta^A(D, \sigma)=\Delta(D, \sigma)$,
\item for any $D \subseteq X \times Q$ and for any $\sigma \in \Sigma_{s, A}$, $\Delta^A(D, \sigma^{\#})=\Delta(D, \sigma)$,
\item $\Delta^A(\varnothing, \$)=D^{\$}$,
\item for any $\varnothing \neq D \subseteq X \times Q$ and for any $\sigma \in \Sigma_{s, A} \cup \Gamma$, $\Delta^A(D, \sigma)=D$. 
\end{enumerate}
 Rule 1) and Rule 2) states that those (and only those) $\sigma \in \Sigma_{s, A}$ transitions of $M$ are relabelled with their attacked copies $\sigma^{\#} \in \Sigma_{s, A}^{\#}$ in  $M^A$. Rule 3) states that it is bad for the attacker if the system execution has been halted and damage has not been inflicted.
Rule 4) is added  so that $M^A$ is over $\Sigma \cup \Sigma_{s, A}^{\#} \cup  \{\$\} \cup \Gamma$ and no event in $\Sigma \cup \Sigma_{s, A}^{\#} \cup  \Gamma$ can be executed at state $\varnothing \in 2^{X \times Q}$. By construction, there is an outgoing transition labelled by each event in $\Sigma \cup \Sigma_{s, A}^{\#} \cup  \Gamma$ at each non-empty state $\varnothing \neq D\subseteq X \times Q$ of $M^A$. 

{\bf Command Execution Automaton}:
The command execution automaton~\cite{LS20},~\cite{LS20J},\\~\cite{zhu2019},~\cite{Linnetworked} is given by the 4-tuple $G_{CE}=(Q^{CE}, \Sigma \cup\Gamma, \delta^{CE},q_0^{CE})$, where $Q^{CE}=\{q^{\gamma} \mid \gamma \in \Gamma\} \cup \{q_{wait}\}$ and  $q_0^{CE}=q_{wait}$. The partial transition function $\delta^{CE}: Q^{CE} \times ( \Sigma \cup\Gamma) \longrightarrow Q^{CE}$ is defined as follows. 
 \begin{enumerate}
     \item for any $\gamma \in \Gamma$, $\delta^{CE}(q_{wait},\gamma)=q^{\gamma}$,
     \item for any $q^\gamma$, if $\sigma \in \Sigma_{uo} \cap \gamma$, $\delta^{CE}(q^{\gamma},\sigma)=q^{\gamma}$,
     \item for any $q^\gamma$, if $\sigma \in \Sigma_{o} \cap \gamma$, $\delta^{CE}(q^{\gamma},\sigma)=q_{wait}$.
 \end{enumerate}
Intuitively, at the initial state $q_{wait}$, the command execution automaton $G_{CE}$ waits
for the supervisor to issue a control command. Once a control command $\gamma$
has been received, it transits to  state $q^{\gamma}$, recording this most recent control command. At state $q^{\gamma}$, only those events
in $\gamma$ are allowed to be fired. If $\sigma \in \Sigma_o$ is fired, then the
command execution automaton returns to the initial state $q_{wait}$ and waits
for a new control command. If $\sigma \in \Sigma_{uo}$ is fired, then temporarily no new control command will be issued by the supervisor and the command execution automaton self-loops $\sigma$ as if $\sigma \in \Sigma_{uo}$ has never occurred. We can view $G_{CE}$ as transducing from control command $\gamma$ issued by the supervisor into event $\sigma \in \Sigma$ executed in the plant.

\begin{figure}[h]
    \centering
    \includegraphics[scale = 0.5]{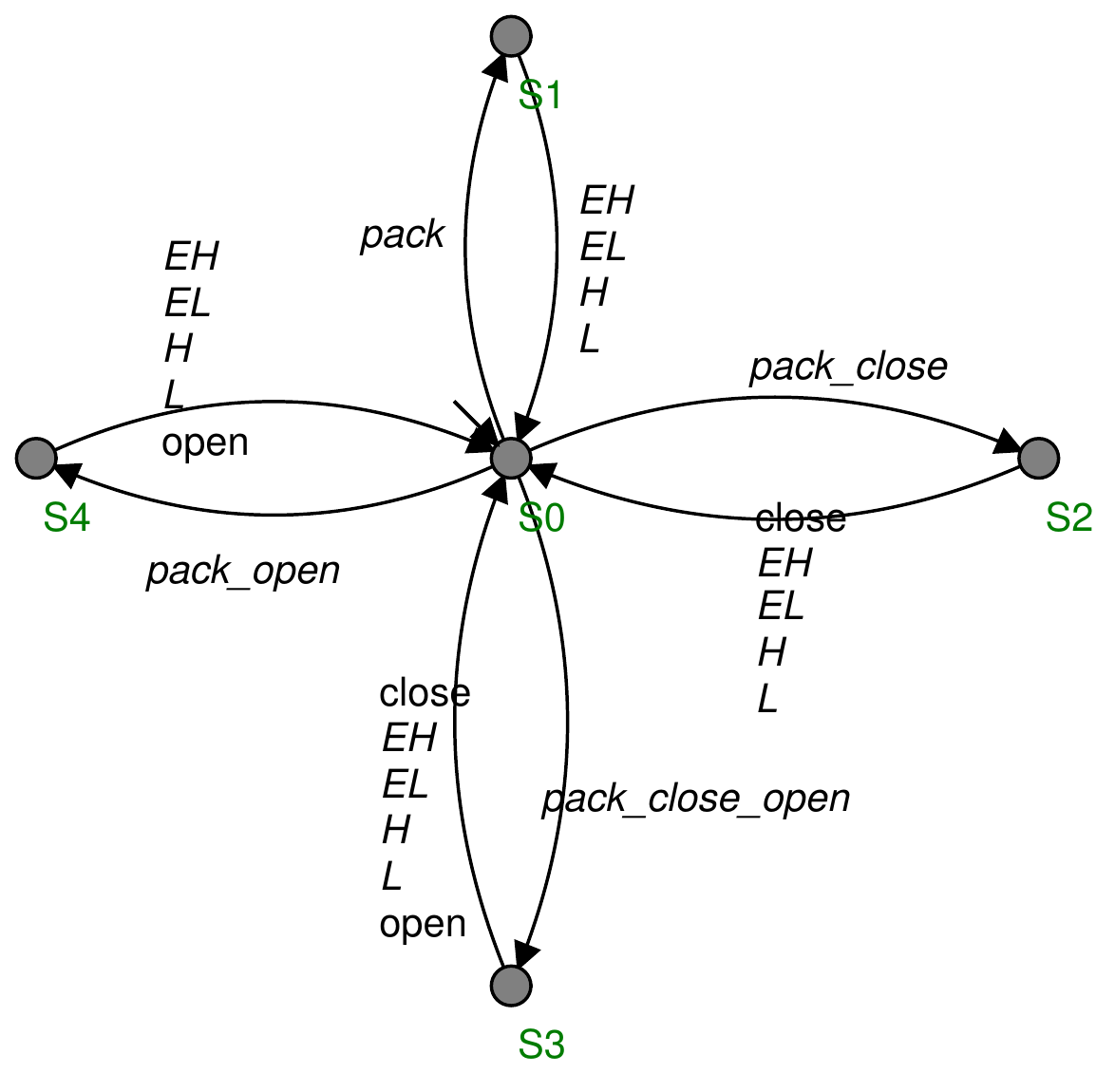}
    \caption{The command execution automaton $G_{CE}$}
    \label{fig:CE}
\end{figure}

\begin{example}
For the water tank example, the command execution automaton $G_{CE}$ is given in Fig.~\ref{fig:CE}.
\end{example}

The event $\$$ in  $G^T\lVert BT(S)^{A} \lVert M^A \lVert G_{SA} \lVert G_{CE}  \lVert A^o$ means that a) the existence of the attacker has been detected by the monitor and the system execution has been halted, and b) damage has not been inflicted, which is different from the meaning of the event $\$$ in  $G^T \lVert M_O^T \lVert G_{SA}\lVert G_{AF} \lVert A^o$. To show the correctness of $A^o$, we need to ensure that $\$$ cannot be executed in the attacked closed-loop system $G^T\lVert BT(S)^{A} \lVert M^A \lVert G_{SA} \lVert G_{CE}  \lVert A^o$, where $\$$ is again uncontrollable and unobservable to the attacker $A^o$, for any supervisor $S$ that is consistent with $O$. 
 \begin{proposition}
No covertness-breaking state  can be reached in the attacked closed-loop system $G^T\lVert BT(S)^{A} \lVert M^A \lVert G_{SA} \lVert G_{CE}  \lVert A^o$, induced by $A^o$, for any supervisor $S$ that is consistent with $O$, i.e., $O \subseteq P_{\Sigma_o}(L(S \lVert G))$. 
\end{proposition}
\noindent {\em Proof}: We need to show that no state in the set $\{(q, x, D, q^{SA}, q^{CE}, y) \in (Q \cup \{q^{\$}\}) \times (X \cup X_{com} \cup \{x^!\}) \times (2^{X \times Q} \cup \{D^{\$}\}) \times Q^{SA} \times Q^{CE} \times Y \mid q \in Q-\{q_{bad}\}, D=\varnothing\}$
can be reached in $G^T\lVert BT(S)^{A} \lVert  M^A \lVert G_{SA} \lVert G_{CE}  \lVert A^o$, so that $\$$ cannot be fired in $G^T\lVert BT(S)^{A} \lVert$\\$  M^A \lVert G_{SA} \lVert G_{CE}  \lVert A^o$.

Suppose some such state $(q, x, D, q^{SA}, q^{CE}, y)$, where $q \in Q-\{q_{bad}\}, D=\varnothing$, can be reached in $G^T\lVert BT(S)^{A} \lVert M^A \lVert 
G_{SA} \lVert G_{CE}  \lVert A^o$ via some string $s \in L(G^T\lVert BT(S)^{A} \lVert M^A \lVert$ \\$G_{SA} \lVert G_{CE}  \lVert A^o)$.
Then, $s$ can be executed in $G^T, M^A$ and $G_{SA}$, after we lift their alphabets to $\Sigma \cup \Sigma_{s, A}^{\#} \cup \{\$\}\cup \Gamma$. By construction, the string $s$ can also be executed in $M_O^T$ and $G_{AF}$, since $s$ can be executed in $M^A$ and $G_{SA}$ and also the event $\$$ does not occur in the string $s$. Thus, $s$ can also be executed in $G^T \lVert M_O^T \lVert G_{SA}\lVert G_{AF} \lVert A^o$. Next, we examine what state is reached in $G^T \lVert M_O^T \lVert G_{SA}\lVert G_{AF} \lVert A^o$ via the string $s$. Clearly, state $q$ is reached in $G^T$ (as in the attacked closed-loop system $G^T\lVert BT(S)^{A} \lVert  M^A \lVert G_{SA} \lVert G_{CE}  \lVert A^o$) and state $u^{!}$ is reached in $M_O^T$ (as $S$ is consistent with $O$) and state $q^{AF, !}$ is reached in $G_{AF}$ (as some sensor replacement attacks have been performed), via the string $s$. It follows that $\$$ can be executed in $G^T \lVert M_O^T \lVert G_{SA}\lVert G_{AF} \lVert A^o$, which is a contradiction to the fact that $A^o$ is a safe attacker for $G^T \lVert M_O^T \lVert G_{SA}\lVert G_{AF}$. We thus can conclude that no  covertness-breaking state  can be reached in $G^T\lVert BT(S)^{A} \lVert M^A \lVert G_{SA} \lVert G_{CE}  \lVert A^o$. \qed \\

Proposition 1 states that $A^o$, synthesized for $G^T \lVert M_O^T \lVert G_{SA}\lVert G_{AF}$, is indeed a covert attacker for the attacked closed-loop system $G^T\lVert BT(S)^{A} \lVert M^A \lVert G_{SA} \lVert G_{CE}\lVert A^o$.  We still need to ensure that $A^o$ is indeed damage-inflicting in $G^T\lVert BT(S)^{A} \lVert M^A \lVert G_{SA} \lVert G_{CE}\lVert A^o$, for any $S$ (and $M$) that is consistent with $O$.  In general, this does not hold, as we have discussed before. But, it can be verified in a straightforward manner, which is shown in the following. Recall that $M_O=(U, \Sigma_o, \eta, u_0)$.
\begin{theorem}
Let $S^{\downarrow}=(U, \Sigma, \eta^S, u_0)$ be a supervisor over the control constraint $(\Sigma_c, \Sigma_o)$, such that $\eta^S$ is the partial transition function defined as follows. 
\begin{enumerate}
    \item for any $u \in U$ and any $\sigma \in \Sigma_o$, $\eta^S(u, \sigma)=\eta(u, \sigma)$,
    \item for any $u \in U$ and any $\sigma \in \Sigma_{uc} \cap \Sigma_{uo}$, $\eta^S(u, \sigma)=u$,
    \item for any $u \in U$ and any $\sigma \in \Sigma_{uc} \cap \Sigma_{o}$, if $\neg \eta(u, \sigma)!$, then  $\eta^S(u, \sigma)=u_{\bot}$.
\end{enumerate}
If $L_m(G^T\lVert BT(S^{\downarrow})^{A} \lVert P_{\Sigma_o}(S^{\downarrow} \lVert G)^A \lVert G_{SA} \lVert G_{CE}\lVert A^o)\neq \varnothing$, then $A^o$ is damage inflicting in $G^T\lVert BT(S)^{A} \lVert M^A \lVert G_{SA} \lVert G_{CE}\lVert A^o$, for any $S$  that is consistent with $O$.
\end{theorem}
\noindent {\em Proof}: $S^{\downarrow}$ is indeed a supervisor over the control constraint $(\Sigma_c, \Sigma_o)$. In particular, by construction,  $S^{\downarrow}$ is an under-approximation of any supervisor $S$ that is consistent with $O$. 
Thus, if damage is reachable with  $S^{\downarrow}$ by $A^o$, i.e., $L_m(G^T\lVert BT(S^{\downarrow})^{A} \lVert P_{\Sigma_o}(S^{\downarrow} \lVert G)^A \lVert G_{SA} \lVert $\\$ G_{CE}\lVert A^o)\neq \varnothing$, then damage is also reachable with any  supervisor $S$ that is consistent with $O$  by $A^o$, since $S$ induces a (non-strictly) larger marked behavior. \qed\\

\begin{figure}[h]
    \centering
    \includegraphics[scale = 0.5]{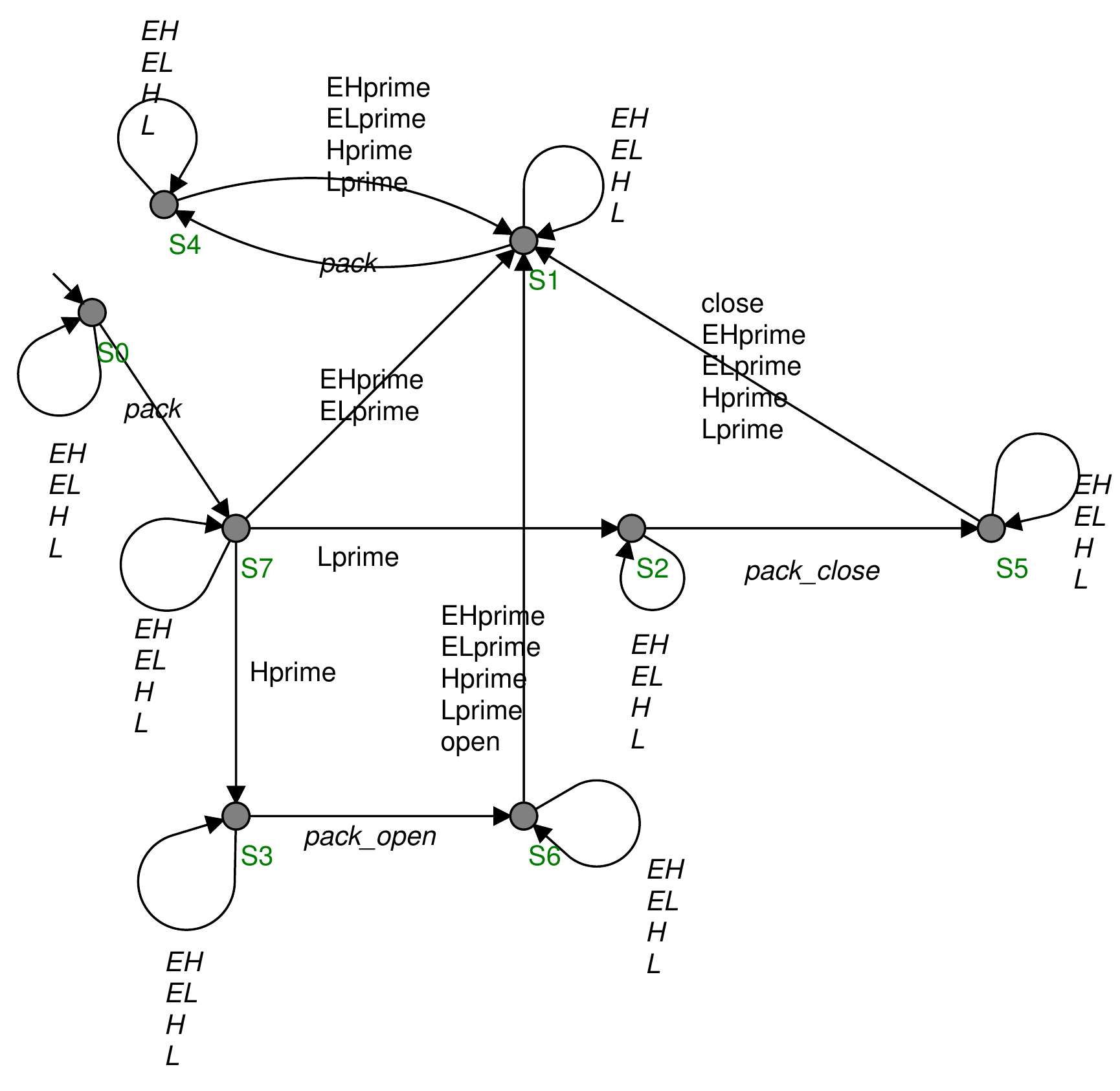}
    \caption{ $BT(S^{\downarrow})^A$ for the water tank example}
    \label{fig:BTSD}
\end{figure}

\begin{example}
For the water tank example, the attacked supervisor $BT(S^{\downarrow})^A$, which is constructed from the supervisor $S^{\downarrow}$ defined in Theorem 1, is given in Fig.~\ref{fig:BTSD}.
\end{example}

We here remark that $S^{\downarrow}$,  defined in Theorem 1,  disables all the events in $\Sigma_c \cap \Sigma_{uo}$. Thus, it introduces some pessimism in the verification of the damage-reachability in general and may result in the rejection of some synthesized covert damage-reachable attacker. That is, the damage-reachability verification of $A^o$ based on $S^{\downarrow}$ in Theorem 1 is sound but in general not complete. However, we shall note that there is a notable exception. $S^{\downarrow}$ is  the least permissive supervisor that is consistent with $O$, not only an under-approximation of any supervisor that is consistent with $O$, when $\Sigma_c \subseteq \Sigma_o$, since we then have $\Sigma_c \cap \Sigma_{uo} =\varnothing$. When $\Sigma_c \subseteq \Sigma_o$, Rule 2) in Theorem 1 becomes: for any $u \in U$ and any $\sigma \in \Sigma_{uo}$, $\eta^S(u, \sigma)=u$.  
\begin{proposition}
$S^{\downarrow}$ is  the least permissive supervisor that is consistent with $O$, when $\Sigma_c \subseteq \Sigma_o$.
\end{proposition}
\noindent {\em Proof}: We first observe the following two facts.
\begin{enumerate}
    \item [i)] $L(S^{\downarrow})=P_{\Sigma_o}^{-1}(O(\Sigma_{uc} \cap \Sigma_o)^*)$, where $P_{\Sigma_o}:\Sigma^* \rightarrow \Sigma_o^*$ .
    \item [ii)] for any supervisor $S$ over the control constraint $(\Sigma_c, \Sigma_o)$, where $\Sigma_c \subseteq \Sigma_o$, we have $L(S)=P_{\Sigma_o}^{-1}(P_{\Sigma_o}(L(S))(\Sigma_{uc} \cap \Sigma_{o})^*)$~\cite{Linthesis}.
\end{enumerate} 

We first show that $S^{\downarrow}$ is consistent with $O$. Since $O$ records a finite set of observations from the closed-loop system $S \lVert G$, we have $O \subseteq P_{\Sigma_o}(L(S \lVert G))$, we then have $O \subseteq P_{\Sigma_o}(L(G))$. Thus, $P_{\Sigma_o}(L(S^{\downarrow}\lVert G))=P_{\Sigma_o}( P_{\Sigma_o}^{-1}(O(\Sigma_{uc} \cap \Sigma_o)^*)  \cap L(G))=P_{\Sigma_o}(L(G)) \cap O(\Sigma_{uc} \cap \Sigma_o)^*\supseteq O$.

Then, we show that $S^{\downarrow}$ is the least permissive supervisor that is consistent with $O$. Let $S$ be any supervisor that is consistent with $O$. 
We have $O \subseteq P_{\Sigma_o}(L(S\lVert G))=P_{\Sigma_o}(L(G) \cap P_{\Sigma_o}^{-1}(P_{\Sigma_o}(L(S))(\Sigma_{uc} \cap \Sigma_o)^*))=P_{\Sigma_o}(L(G)) \cap P_{\Sigma_o}(L(S))(\Sigma_{uc}\cap \Sigma_o)^*$. We have $O \subseteq P_{\Sigma_o}(L(S))(\Sigma_{uc}\cap \Sigma_o)^*$ and  $L(S^{\downarrow})=P_{\Sigma_o}^{-1}(O(\Sigma_{uc} \cap \Sigma_o)^*) \subseteq P_{\Sigma_o}^{-1}(P_{\Sigma_o}(L(S))(\Sigma_{uc} \cap \Sigma_{o})^*) = L(S)$. \qed \\

We have the following characterization of the damage-reachability of $A^o$ when $\Sigma_c \subseteq \Sigma_o$.
\begin{theorem}
Suppose $\Sigma_c \subseteq \Sigma_o$, then $L_m(G^T\lVert BT(S^{\downarrow})^{A} \lVert P_{\Sigma_o}(S^{\downarrow} \lVert G)^A \lVert G_{SA} \lVert G_{CE}\lVert A^o)\neq \varnothing$ iff $A^o$ is damage inflicting in $G^T\lVert BT(S)^{A} \lVert M^A \lVert G_{SA} \lVert G_{CE}\lVert A^o$, for any $S$  that is consistent with $O$.
\end{theorem}
\noindent {\em Proof}: This immediately follows from Theorem 1 and the fact that $S^{\downarrow}$ is  the least permissive supervisor that is consistent with $O$ when $\Sigma_c \subseteq \Sigma_o$. \qed\\

Theorem 1 is quite  straightforward. But, there are two mild disadvantages. First of all, it involves the construction of $P_{\Sigma_o}(S^{\downarrow} \lVert G)^A$, which requires a subset construction. Secondly, most of the existing tools directly output $G^T \lVert M_O^T \lVert G_{SA}\lVert G_{AF} \lVert A^o$, instead of $A^o$ which requires another extraction step~\cite{B1993}. In the following, we show how we can directly use $G^T \lVert M_O^T \lVert G_{SA}\lVert G_{AF} \lVert A^o$ for the verification. We have the following useful result.
\begin{proposition}
The following equality holds:
\begin{center}
$L_m(G^T \lVert M_O^T \lVert G_{SA}\lVert G_{AF} \lVert A^o \lVert  BT(S^{\downarrow})^{A} \lVert G_{CE})=L_m(G^T\lVert BT(S^{\downarrow})^{A} \lVert P_{\Sigma_o}(S^{\downarrow} \lVert G)^A \lVert G_{SA} \lVert G_{CE}\lVert A^o)$
\end{center}
\end{proposition}
\noindent {\em Proof}:
($\subseteq$): Let $s$ be any string in $L_m(G^T \lVert M_O^T \lVert G_{SA}\lVert G_{AF} \lVert A^o \lVert BT(S^{\downarrow})^{A} \lVert G_{CE})$. It holds that $s$ can be executed in $G^T, BT(S^{\downarrow})^{A}, G_{SA}, G_{CE}, A^o, M_O^T, G_{AF}$, after we lift their alphabets to $\Sigma \cup \Sigma_{s, A}^{\#} \cup \{\$\} \cup \Gamma$. 

We first observe that the $q_{bad}$ state is reached in $G^T$, some state $u \in U \cup \{u^{!}\}$ is reached in $M_O^T$ and the state $q^{AF, !}$ is reached in $G_{AF}$, via string $s$. $s$ must be of the form  $s=s_1\sigma$, for some $s_1 \in (\Sigma \cup \Sigma_{s, A}^{\#} \cup \Gamma)^*$ and some $\sigma \in \Sigma$, such that the $q_{bad}$ state is reached in $G^T$ via the string $s_1\sigma$ and the $q^{AF, !}$ state is reached in $G_{AF}$ via $s_1$. Thus, the $q_{bad}$ state has not been reached in $G^T$ via the string $s_1$. We conclude that some state in $U $ is reached in $M_O^T$ via $s_1$; otherwise $s_1\$$ can be executed in $G^T \lVert M_O^T \lVert G_{SA}\lVert G_{AF} \lVert A^o$, which is impossible. Thus, we conclude that some state $D \in 2^{X \times Q}-\{\varnothing\}$ is reached in    $P_{\Sigma_o}(S^{\downarrow} \lVert G)^A$ via $s_1$ and some state $D' \in 2^{X \times Q}$ is reached in $P_{\Sigma_o}(S^{\downarrow} \lVert G)^A$ via $s=s_1\sigma$. That is, $s$ can be executed in $P_{\Sigma_o}(S^{\downarrow} \lVert G)^A$. 

It then follows that $s$ can be executed in $G^T\lVert BT(S^{\downarrow})^{A} \lVert P_{\Sigma_o}(S^{\downarrow} \lVert G)^A \lVert G_{SA} \lVert G_{CE}\lVert A^o$ and a marked state of $G^T\lVert BT(S^{\downarrow})^{A} \lVert P_{\Sigma_o}(S^{\downarrow} \lVert G)^A \lVert G_{SA} \lVert G_{CE}\lVert A^o$ is indeed reached via the string $s$. We can then conclude that $s \in L_m(G^T\lVert BT(S^{\downarrow})^{A} \lVert P_{\Sigma_o}(S^{\downarrow} \lVert G)^A \lVert G_{SA} \lVert G_{CE}\lVert $\\$ A^o)$.

($\supseteq$): $G^T \lVert M_O^T \lVert G_{SA}\lVert G_{AF} \lVert A^o$  by construction provides  a (behavioral) upper-bound for $G^T\lVert BT(S^{\downarrow})^{A} \lVert P_{\Sigma_o}(S^{\downarrow} \lVert G)^A \lVert G_{SA} \lVert G_{CE}\lVert A^o$. Thus, we have
\begin{center}$L(G^T\lVert BT(S^{\downarrow})^{A} \lVert P_{\Sigma_o}(S^{\downarrow} \lVert G)^A \lVert G_{SA} \lVert G_{CE}\lVert A^o) \subseteq L(G^T \lVert M_O^T \lVert G_{SA}\lVert G_{AF} \lVert A^o)$ \end{center} and 
\begin{center}
$L_m(G^T\lVert BT(S^{\downarrow})^{A} \lVert P_{\Sigma_o}(S^{\downarrow} \lVert G)^A \lVert G_{SA} \lVert G_{CE}\lVert A^o) \subseteq L_m(G^T \lVert M_O^T \lVert G_{SA}\lVert G_{AF} \lVert A^o)$.
\end{center} 
To see the first inclusion, we first remark that $\$$ cannot be executed in $G^T\lVert BT(S^{\downarrow})^{A} \lVert$\\$ P_{\Sigma_o}(S^{\downarrow} \lVert G)^A \lVert G_{SA} \lVert G_{CE} \lVert A^o$ or $G^T \lVert M_O^T \lVert G_{SA}\lVert G_{AF} \lVert A^o$.
We only need to show that for each component in the right hand side, it is lower bounded by some component in the left hand side, when their alphabets are lifted to $\Sigma \cup \Sigma_{s, A}^{\#}\cup \{\$\} \cup \Gamma$. For $M_O^T$, it is  lower bounded by $P_{\Sigma_o}(S^{\downarrow} \lVert G)^A$; for $G_{AF}$, it is lower bounded by $G_{SA}$, as $\$$ cannot be executed in $G_{AF}$ or $G_{SA}$. 

To see the second inclusion, let $s \in L_m(G^T\lVert BT(S^{\downarrow})^{A} \lVert P_{\Sigma_o}(S^{\downarrow} \lVert G)^A \lVert G_{SA} \lVert G_{CE}\lVert A^o)$. We know that $s \in L(G^T \lVert M_O^T \lVert G_{SA}\lVert G_{AF} \lVert A^o)$. It is clear that the state $q_{bad}$  is reached in $G^T$, some state $u \in U \cup \{u!\}$ is reached in $M_O^T$, the state $q^{AF, !}$ is reached in $G_{AF}$. Thus, $s \in L_m(G^T \lVert M_O^T \lVert G_{SA}\lVert G_{AF} \lVert A^o)$.
\qed\\

With Theorem 1, Theorem 2 and Proposition 2, the following result is then immediate. It says that we only need to verify
$L_m(G^T \lVert M_O^T \lVert G_{SA}\lVert G_{AF} \lVert A^o \lVert  BT(S^{\downarrow})^{A} \lVert G_{CE})\neq \varnothing$, instead of verifying $L_m(G^T\lVert BT(S^{\downarrow})^{A} \lVert P_{\Sigma_o}(S^{\downarrow} \lVert G)^A \lVert G_{SA} \lVert G_{CE}\lVert A^o)\neq \varnothing$.
\begin{corollary}
If $L_m(G^T \lVert M_O^T \lVert G_{SA}\lVert G_{AF} \lVert A^o \lVert  BT(S^{\downarrow})^{A} \lVert G_{CE})\neq \varnothing$, then $A^o$ is damage inflicting in $G^T\lVert BT(S)^{A} \lVert M^A \lVert G_{SA} \lVert G_{CE}\lVert A^o$, for any $S$  that is consistent with $O$. Suppose $\Sigma_c \subseteq \Sigma_o$, then $L_m(G^T \lVert M_O^T \lVert G_{SA}\lVert G_{AF} \lVert A^o \lVert  BT(S^{\downarrow})^{A} \lVert G_{CE})\neq \varnothing$ iff $A^o$ is damage inflicting in $G^T\lVert BT(S)^{A} \lVert M^A \lVert G_{SA} \lVert G_{CE}\lVert A^o$, for any $S$  that is consistent with $O$.
\end{corollary}

Intuitively, the synthesized attacker $A^o$ can ensure the damage-reachability in the attacked closed-loop system, for any supervisor $S$ that is consistent with $O$, if there exists a marked string $s \in L_m(G^T \lVert M_O^T \lVert G_{SA}\lVert G_{AF} \lVert  A^o)$ such that $s$ is allowed by an attacked supervisor $S^{\downarrow, A}$ where $S^{\downarrow}$ under-approximates any supervisor that is consistent with $O$, i.e., $s$ can be executed in $BT(S^{\downarrow})^{A} \lVert G_{CE}$. 

\begin{figure}[h]
    \centering
    \includegraphics[scale = 0.44]{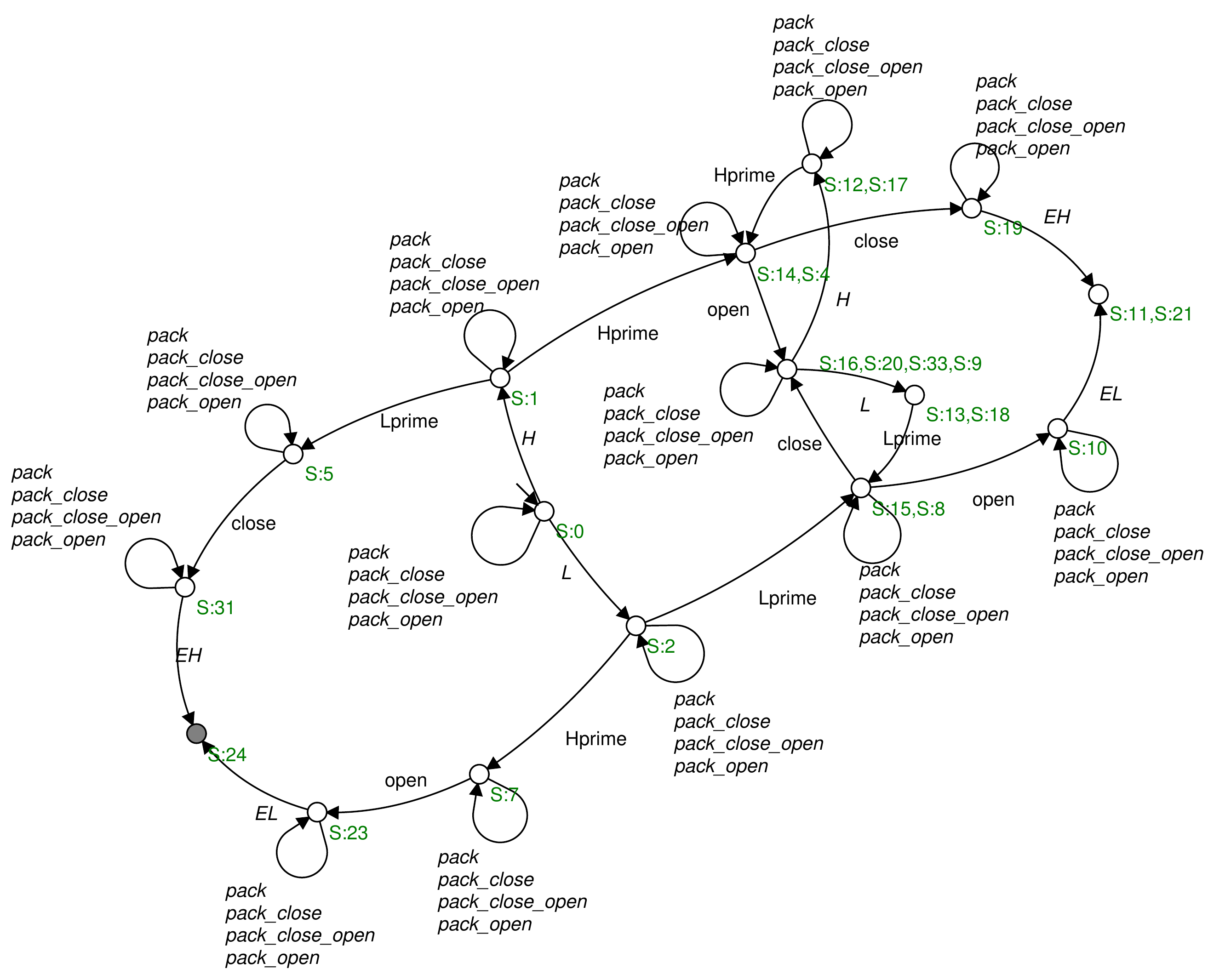}
    \caption{The synthesized safe attacker based on the surrogate plant}
    \label{fig:attacker}
\end{figure}

\begin{figure}[h]
    \centering
    \includegraphics[scale = 0.45]{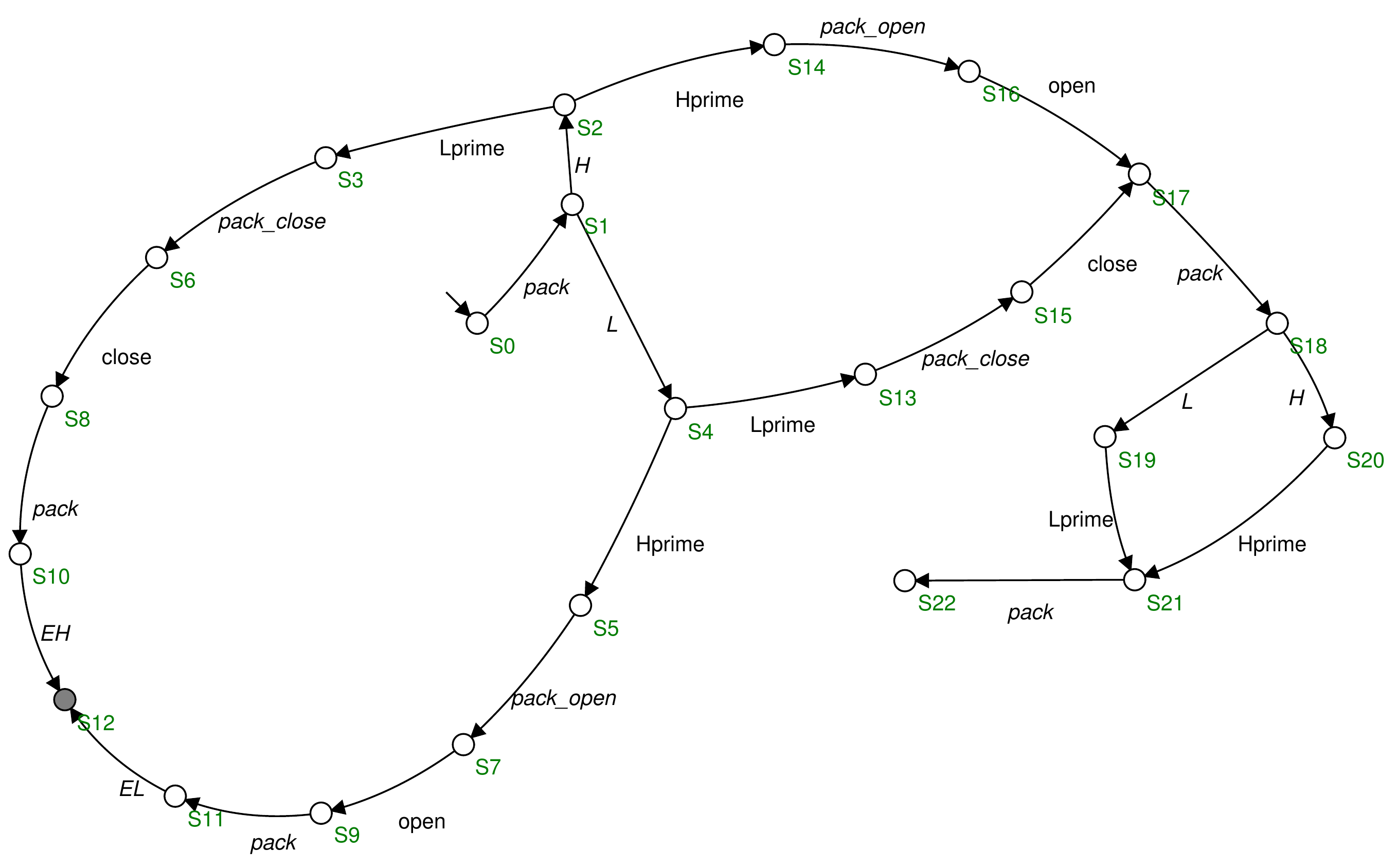}
    \caption{The damage-reachability verification on the synthesized covert attacker}
    \label{fig:verify}
\end{figure}

\begin{example}
For the water tank example, a non-empty safe attacker has been synthesized by using the Supremica~\cite{Malik07}, which is given in the form of $G^T \lVert M_O^T \lVert G_{SA}\lVert G_{AF} \lVert A^o$ and shown in Fig.~\ref{fig:attacker}. Then, by applying Corollary 1, we can compute $G^T \lVert M_O^T \lVert G_{SA}\lVert $\\$ G_{AF} \lVert A^o \lVert  BT(S^{\downarrow})^{A} \lVert G_{CE}$, which is shown in Fig.~\ref{fig:verify}. With a reachable marked state, i.e., $S12$, Fig.~\ref{fig:verify} confirms that the synthesized attacker in Fig.~\ref{fig:attacker} is indeed damage-reachable, in addition to being covert, against all the supervisors that are consistent with $O$. 
\end{example}

We now briefly discuss on the complexity of the above approach.

The size of the reachable state space of the surrogate plant $G^T \lVert M_O^T \lVert G_{SA}\lVert G_{AF}$ is no more than $Z=(|Q|+1)(|U|+2)(|\Sigma_{s, A}+1|)(|\Sigma_{s, A}|+3)$. Depending on the choices of the adopted supervisor synthesis algorithms (see, for example,~\cite{WMW10},~\cite{YL16},~\cite{WLLW18}), the complexity for the synthesis of non-empty safe attacker $A^o$ could differ. For example, suppose we adopt the normality property based\footnote{Since the control constraint is $(\Sigma_{a, A} \cup \Sigma_{s, A}^{\#}, \Sigma_{o}  \cup \Sigma_{s, A}^{\#})$ (see Section~\ref{sec: SS}), the normality property based synthesis approach effectively only allows the attacker to control those events in $(\Sigma_{a, A} \cap \Sigma_o) \cup \Sigma_{s, A}^{\#}$.} synthesis approach~\cite{WMW10},~\cite{WLLW18}, the time complexity for the synthesis is $\mathcal{O}((|\Sigma|+|\Sigma_{s, A}| + |\Gamma|)2^Z)$, since we do not require the nonblockingness. To perform the damage-reachability verification, we need to verify
\begin{center}
$L_m(G^T \lVert M_O^T \lVert G_{SA}\lVert G_{AF} \lVert A^o \lVert  BT(S^{\downarrow})^{A} \lVert G_{CE})\neq \varnothing.$
\end{center}
The state size of $G^T \lVert M_O^T \lVert G_{SA}\lVert G_{AF} \lVert A^o \lVert  BT(S^{\downarrow})^{A} \lVert G_{CE}$ is no more than $Z2^Z(2|U|+1)(|\Gamma|+1)$. Thus, the time complexity for  damage-reachability verification is $\mathcal{O}((|\Sigma|+|\Sigma_{s, A}| + |\Gamma|)|U||\Gamma|Z2^Z)$. Thus, the total time complexity is $\mathcal{O}((|\Sigma|+ |\Gamma|)|U||\Gamma|Z2^Z)$, where $Z=(|Q|+1)(|U|+2)(|\Sigma_{s, A}+1|)(|\Sigma_{s, A}|+3)$.

There remains one  disadvantage of the above approach, that is, it requires a synthesis step followed by a verification step. Indeed, if the synthesized (maximally permissive) covert attacker $A^o$ is verified to be not damage-reachable, then (in general) we may need to keep searching for another (maximal permissive) covert attacker and hope it is indeed damage-reachable. This is troublesome. In the following, we show how the synthesis can be carried out in such a way that damage-reachability is always ensured by construction. The key insight is to embed the verification step within the synthesis step. 

To that end, we need to bring in the concept of automaton completion~\cite{LS20BJ}. Formally, the completion of any (partial) finite-state automaton $P=(I, \Sigma, \pi, i_0, I_m)$ is a complete finite-state automaton $\overline{P}=(I \cup \{i_d\}, \Sigma, \overline{\pi}, i_0, I_m)$, where the distinguished state $i_d \notin I$ denotes the added dump state and  $\overline{\pi}=$
\begin{center}
$\pi \cup (\{i_d\} \times \Sigma \times \{i_d\}) \cup \{(i, \sigma, i_d) \mid \pi(i, \sigma)$ is undefined$, i \in I, \sigma \in \Sigma\}$
\end{center}
denotes the transition function. 

Let $\overline{BT(S^{\downarrow})^{A}}$ denote the completion of   $BT(S^{\downarrow})^{A}$ and let $\overline{G_{CE}}$ denote the completion of $G_{CE}$. We now construct the  surrogate plant $G^T \lVert M_O^T \lVert G_{SA}\lVert G_{AF} \lVert \overline{BT(S^{\downarrow})^{A}} \lVert \overline{G_{CE}}$ and perform the synthesis of (maximally permissive) safe attacker on $G^T \lVert M_O^T \lVert G_{SA}\lVert $\\$G_{AF} \lVert  \overline{BT(S^{\downarrow})^{A}} \lVert \overline{G_{CE}}$ instead; as before, the set of bad states are again specified by (the combination of) the state $q^{\$}$ in $G^T$, the state $u^{\$}$ in $M_O^T$ and the state $q^{AF,\$}$ in $G_{AF}$. 
We immediately have the following useful result.
\begin{theorem}
Any non-empty safe attacker for  $G^T \lVert M_O^T \lVert G_{SA}\lVert G_{AF} \lVert  \overline{BT(S^{\downarrow})^{A}} \lVert \overline{G_{CE}}$ is a covert damage-reachable attacker on the attacked closed-loop system $G^T\lVert BT(S)^{A} \lVert $\\$M^A \lVert G_{SA} \lVert G_{CE}  \lVert A$ induced by $A$, for any supervisor $S$ that is consistent with $O$, i.e., $O \subseteq P_{\Sigma_o}(L(S \lVert G))$.
\end{theorem}
\noindent {\em Proof}:
We first observe that $L(G^T \lVert M_O^T \lVert G_{SA}\lVert G_{AF} \lVert  \overline{BT(S^{\downarrow})^{A}} \lVert \overline{G_{CE}})=L(G^T \lVert M_O^T \lVert G_{SA}\lVert $\\$ G_{AF})$ and also the sets of bad strings for $G^T \lVert M_O^T \lVert G_{SA}\lVert G_{AF} \lVert  \overline{BT(S^{\downarrow})^{A}} \lVert \overline{G_{CE}}$ and $G^T \lVert M_O^T $\\$\lVert G_{SA}\lVert  G_{AF}$ are the same. Thus, any safe attacker for $G^T \lVert M_O^T \lVert G_{SA}\lVert G_{AF} \lVert  \overline{BT(S^{\downarrow})^{A}} \lVert \overline{G_{CE}}$ is also a safe attacker for $G^T \lVert M_O^T \lVert G_{SA}\lVert  G_{AF}$, and vice versa. 

Let $A^o$ be any  non-empty safe attacker for $G^T \lVert M_O^T \lVert G_{SA}\lVert G_{AF} \lVert  \overline{BT(S^{\downarrow})^{A}} \lVert \overline{G_{CE}}$. By the above analysis, we know that $A^o$ is a safe attacker for $G^T \lVert M_O^T \lVert G_{SA}\lVert  G_{AF}$. Furthermore, we know that there exists some string $s \in L_m(G^T \lVert M_O^T \lVert G_{SA}\lVert G_{AF} \lVert A^o \lVert \overline{BT(S^{\downarrow})^{A}} \lVert $\\$ \overline{G_{CE}})=L_m(G^T \lVert M_O^T \lVert G_{SA}\lVert G_{AF} \lVert A^o \lVert BT(S^{\downarrow})^{A} \lVert G_{CE})$. Thus, by Proposition 1 and Corollary 1, we know that $A^o$ is a covert damage-reachable attacker on the attacked closed-loop system $G^T\lVert BT(S)^{A} \lVert M^A \lVert G_{SA} \lVert G_{CE}  \lVert A$ induced by $A$, for any supervisor $S$ that is consistent with $O$, i.e., $O \subseteq P_{\Sigma_o}(L(S \lVert G))$. \qed \\ 

Intuitively, we only need to synthesize a non-empty safe attacker $A^o$ for $G^T \lVert M_O^T \lVert $\\$ G_{SA}\lVert G_{AF} \lVert  \overline{BT(S^{\downarrow})^{A}} \lVert \overline{G_{CE}}$; and $A^o$ is guaranteed to be covert damage-reachable on the attacked closed-loop system induced by any supervisor $S$ that is consistent with $O$. Due to our use of over-approximation, the other direction of the implication in general does not hold for Theorem 3. 

Next, we briefly discuss on the complexity of the approach based on Theorem 3.

The state size of $G^T \lVert M_O^T \lVert G_{SA}\lVert G_{AF} \lVert  \overline{BT(S^{\downarrow})^{A}} \lVert \overline{G_{CE}}$ is $Z(2|U|+2)(|\Gamma|+2)$. Thus, the time complexity of the synthesis approach by using Theorem 3, if we adopt the normality property based synthesis approach, costs $\mathcal{O}((|\Sigma|+|\Gamma|)2^{Z(2|U|+2)(|\Gamma|+2)})$. 
\section{Conclusions and Future Works}
This paper studies the covert attacker synthesis problem in the new setup where the model of the supervisor is unknown, but a (prefix-closed) finite set of observations of the runs of the closed-loop system is assumed to be available, to the adversary. We have proposed a heuristic synthesis algorithm, with a formal guarantee of correctness, for the synthesis of covert damage-reachable attackers,  even without using the model of the supervisor. The solution methodology is to formulate the observation-assisted covert attacker synthesis problem as an instance of the partial-observation supervisor synthesis problem. It then follows that we can use the existing supervisor synthesis solvers to compute non-empty (i.e., damage-reachable) safe (i.e., covert) attackers.

There are several limitations of this work that need to be addressed. First of all, we only consider sensor replacement attacks and actuator disablement attacks; it is of interest to consider other attack mechanisms. Our approach seems capable of dealing with sensor insertion and deletion attacks, without much modification, but it has some difficulty in coping with actuator enablement attacks. Secondly, the synthesized covert damage-reachable attacker is in general not maximally permissive, due to the use of  over-approximation in the surrogate plant. Thus, the approach has not resolved the  decidability of the observation-assisted covert  attacker synthesis problem.  
Lastly, we have not addressed the problem of synthesis of covert damage-nonblocking attackers, which seems to be more challenging. These and other issues are to be addressed in the future works. 

{\bf Acknowledgements}: The research of the project was supported by Ministry of Education, Singapore, under grant AcRF TIER 1-2018-T1-001-245 (RG 91/18) and supported by the funding from Singapore National Research Foundation via Delta-NTU Corporate Lab Program (DELTA-NTU CORP
LAB-SMA-RP2 SU RONG M40\\61925.043). We would like to thank the anonymous reviewers for comments that help us improve the quality of the paper.


\begin{thebibliography}{}

\bibitem{WMW10} 
W. M. Wonham, K. Cai, \emph{Supervisory control of discrete-event
systems}, Monograph Series Communications and Control Engineering,
Springer, 2018.

\bibitem{CarvalhoEnablementAttacks} 
L. K. Carvalho, Y. C. Wu, R. Kwong, S. Lafortune, {``Detection and prevention of actuator enablement attacks in supervisory control systems"}, International Workshop on Discrete Event Systems, pp. 298-305, 2016.

\bibitem{Carvalho2018}
L. K. Carvalho,  Y. C. Wu,  R. Kwong, S.  Lafortune,  {``Detection and  mitigation of  classes of attacks  in supervisory  control systems"}, Automatica, vol. 97, pp. 121-133, 2018.

\bibitem{LACM17}
P. M. Lima, M. V. S. Alves, L. K. Carvalho, M. V. Moreira, {``Security against network attacks in supervisory control systems"}, IFAC, 50(1): 12333-12338, 2017

\bibitem{Lima2018}
P. M. Lima, L. K. Carvalho, M. V.  Moreira,  {``Detectable  and undetectable network attack security of cyber-physical systems"}, IFAC, 51(7): 179-185, 2018.

\bibitem{WTH17}
M. Wakaiki, P. Tabuada, J. P. Hespanha, {``Supervisory control of discrete-event systems under attacks"}, Dynamic Games and Applications, 9: 965–983, 2019.

\bibitem{WP}
Y. Wang, M. Pajic, 
{``Supervisory control of discrete event systems in the presence of sensor and actuator attacks"}, Conference on Decision and Control, pp. 5350-5355, 2019.



\bibitem{Goes2017}
R. Meira-Goes, E. Kang, R. Kwong, S. Lafortune, {``Stealthy deception attacks for cyber-physical systems"}, Conference on Decision and Control, pp: 4224-4230, 2017. 

\bibitem{Goes2020}
R. Meira-Goes, E. Kang, R. Kwong, S. Lafortune, {``Synthesis of sensor deception attacks at the supervisory layer of cyber–physical systems"}, Automatica, vol. 121, 109172, 2020.

\bibitem{Mohajerani20}
S. Mohajerani,  R. Meira-Goes,
S. Lafortune, {``Efficient synthesis of sensor deception attacks using observation
equivalence-based abstraction"},  Workshop on Discrete Event Systems, pp: 28-34, 2020.


\bibitem{Lin2018}
L. Lin, S.  Thuijsman, Y.  Zhu, S. Ware, R.  Su, M. Reniers, {``Synthesis of  successful  actuator  attackers on  supervisors"}, American Control Conference, pp: 5614-5619, 2019.

\bibitem{LZS19}
L. Lin, Y. Zhu, R. Su, {``Synthesis of covert actuator attackers for free"}, Discrete Event Dynamic Systems:    Theory and Applications, DOI: 10.1007/s10626-020-00312-2, 2020.

\bibitem{LS20}
L. Lin, R. Su, {``Synthesis of covert actuator and sensor attackers as supervisor synthesis"}, Workshop on Discrete Event Systems, pp: 1-6, 2020.

\bibitem{LS20J}
L. Lin, R. Su, {``Synthesis of covert actuator and sensor attackers"}, Automatica, vol 130, 109714, 2021.

\bibitem{Kh19}
A. Khoumsi, {``Sensor and actuator attacks of cyber-physical systems: a study based on supervisory control of discrete event systems"},  Conference on Systems and Control, pp. 176-182, 2019.


\bibitem{Su2018}
R. Su, {``Supervisor synthesis to thwart cyber-attack with bounded sensor reading alterations"}, Automatica, vol 94, pp.35-44, 2018.

\bibitem{Su20}
R. Su, {``On decidability of existence of nonblocking supervisors
resilient to smart sensor attacks"}, arXiv:2009.02626v1, 2020.

\bibitem{GSS19}
R. Meira-Goes, H. Marchand, S. Lafortune {``Towards resilient supervisors against sensor deception attacks"},  Conference on Decision and Control, pp. 5144-5149, 2019.

\bibitem{WBP19}
Y.~Wang, M. Pajic, 
{``Attack-resilient supervisory control with intermittently secure communication"},  Conference on Decision and Control, pp: 2015-2020, 2019.



\bibitem{Zhu2018}
Y. Zhu, L. Lin, R. Su, {``Supervisor obfuscation against actuator enablement attack"}, European Control Conference, pp: 1760-1765, 2019.

\bibitem{LZS19b}
L. Lin, Y. Zhu, R. Su, {``Towards bounded synthesis of resilient supervisors"},  Conference on Decision and Control, pp: 7659-7664, 2019.

\bibitem{LS20BJ}
L. Lin, R. Su, {``Bounded synthesis of resilient supervisors"},  IEEE Transactions on Automatic Control, under review, 2020.


\bibitem{Susyna}
SuSyNA: Supervisor synthesis for non-deterministic automata. 2011 [Online]. Available:https://www.ntu.edu.sg/home/rsu/Downloads.htm.

\bibitem{Feng06}
L. Feng, W. M. Wonham, {``Tct: A computationtool for supervisory control synthesis"},   Workshop on Discrete Event Systems, pages 388–389, 2006.

\bibitem{Malik07}
R. Malik, K. Akesson, H. Flordal, M. Fabian, {``Supremica--an efficient tool for large-scale discrete event systems"}, IFAC-PapersOnLine,  50: 5794–5799, 2017.


















\bibitem{CL99}
C. Cassandras, S. Lafortune, \emph{Introduction to discrete event systems}, Boston, MA: Kluwer, 1999.

\bibitem{HU79}
J.~E. Hopcroft, J.~D. Ullman, \emph{Introduction to automata theory, languages, and computation}, Addison-Wesley, Reading, Massachusetts, 1979.
















 
 
 



\bibitem{B1993}
A. Bergeron, {``A unified approach to control problems in discrete event processes"},
RAIRO-Theoretical Informatics and Applications, 27(6): 555-573,
1993.



\bibitem{zhu2019}
Y. Zhu, L. Lin, S. Ware, R. Su, {``Supervisor synthesis for networked discrete event systems with communication delays and lossy events"}, Conference on Decision and Control, pp. 6730-6735, 2019.

\bibitem{Linnetworked}
L. Lin, Y. Zhu, R. Tai, S. Ware, R. Su, {``Networked supervisor synthesis against lossy  channels with bounded network delays as non-networked synthesis "}, Automatica, under review, 2020.

\bibitem{Linthesis}
L. Lin, {``Towards decentralized and parameterized supervisor synthesis"}, Ph.D thesis, School of Electrical and Electronic Engineering, Nanyang Technological Unversity, 2015. [Online] Available: https://dr.ntu.edu.sg/handle/10356/65641.

\bibitem{YL16}
X. Yin, S. Lafortune,  
{``Synthesis of maximally permissive supervisors for partially observed discrete event systems"}, IEEE Transactions on Automatic Control, 61(5):1239-1254, 2016.

\bibitem{WLLW18}
D. Wang, L. Lin, Z. Li, W. M. Wonham, {``State-based control of discrete-event systems under partial observation"}, IEEE Access, 6(1): 42084-42093, 2018. 

\end{thebibliography}
\end{document}